\documentclass[pre,twocolumn,aps,superscriptaddress,showpacs,floatfix]{revtex4}
\usepackage{graphicx}
\usepackage{dcolumn}
\usepackage{bm}
\usepackage{amsmath}
\usepackage{amssymb}

\begin{document}
\title{Ratcheting of driven attracting colloidal particles: Temporal density oscillations and current multiplicity}  
\author{Andrey Pototsky}
\affiliation{Department of Mathematics, University of Cape Town,
Rondebosch 7701, South Africa}
\author{Andrew J. Archer}
\affiliation{Department of Mathematical Sciences, Loughborough University,
Loughborough LE11 3TU, United Kingdom}
\author{Sergey E. Savel'ev}
\affiliation{Department of Physics, Loughborough University,
Loughborough LE11 3TU, United Kingdom}
\author{Uwe Thiele}
\affiliation{Department of Mathematical Sciences, Loughborough University,
Loughborough LE11 3TU, United Kingdom}
\author{Fabio Marchesoni}
\affiliation{Dipartimento di Fisica, Universit\`a di Camerino,
I-62032 Camerino, Italy}

\begin{abstract}
We consider the unidirectional particle transport in a suspension of colloidal particles which interact with each other via a pair potential having a hard-core repulsion plus an attractive tail. The colloids are confined within a long narrow channel and are driven along by a DC or an AC external potential. In addition, the walls of the channel interact with the particles via a ratchet-like periodic potential. We use dynamical density functional theory to compute the average particle current. In the case of DC drive, we show that as the attraction strength between the colloids is increased beyond a critical value, the stationary density distribution of the particles loses its stability leading to depinning and a time dependent density profile. Attraction induced symmetry breaking gives rise to the coexistence of stable stationary density profiles with different spatial periods and time-periodic density profiles, each characterized by different values for the particle current.
\end{abstract}

 \maketitle

\section{Introduction}
Much attention has been given to studying the transport of particles along narrow channels \cite{RMP09}. Such strong confinement occurs for example in the ion channels of biological membranes \cite{hille01}, in zeolites and other porous materials \cite{kaerger92} and in microfluidic devices \cite{squires05}. Experimental studies of colloidal particles confined within grooves etched on a surface \cite{wei00} have already addressed the case when confinement is so extreme that  particles cannot pass one another and so single-file diffusion sets in \cite{lutz04}. In many cases, the motion of a file occurs in the presence of a periodic pinning potential. The latter can be induced, for instance, by defects, such as in the case of superconducting vortices moving in easy-flow channels \cite{besseling98}, by other particles, as in the case of a fluctuating quasi-1-dimensional (1D) channel \cite{coupier07}, by a periodic distribution of charges, as in the case of the motor proteins moving along a microtubulus \cite{ashkin90}, or, more generally, by a periodic corrugation of the channel walls \cite{wambaugh,taloni06}. When the left-right symmetry of the pinning potential is broken, an externally applied center-symmetric AC drive induces a net drift of the file in a certain direction. The efficiency of such a rectification mechanism strongly depends on the number of particles in the file, their size, and frequency of the drive \cite{vicsek95}.

In addition to the interaction with the channel, the colloids have an excluded volume interaction between them \cite{BarratHansen2003, vicsek95} and may also exhibit mutual attraction, either due to Van der Waals forces \cite{hansen2006tsl} or because of the presence of other passive molecules in the solution, such as in the case of colloid-polymer mixtures \cite{poon02, stradner04, savsav}. It has been recognized that attractive forces lead to the formation of particle clusters and, consequently, to a dramatic increase in particle diffusion and mobility \cite{sholl97}. Such enhancement is explained by the mismatch between the size of the particle clusters and the characteristic length scale of the corrugated potential induced by the walls of the channel. In the case of the diffusion of long alkane chains in zeolites, a similar phenomenon is called the ``window effect" \cite{dubbeldam03}, namely, the mobility of the alkane chain becomes enhanced, whenever its length is not commensurate with the zeolite cage. More generally, the incommensurability between the lateral dimensions of a biological molecule and the size of a catalyst is known as ``shape selectivity" \cite{smit08}, a recurrent scheme utilized by nature to control enzymatic reactions in living cells.

Recently, we have developed a theory \cite{PA10}, based on dynamical density functional theory (DDFT) \cite{marini99, marini00, archer04, archer04b}, that captures the essential features of the condensation process from the disordered to the condensed state in a randomly distributed single-file of interacting particles. Pair attraction can be used to enhance the transport of colloidal particles in 1D. For instance, entrained attracting particles can be effectively shuttled along an asymmetric corrugated channel by means of a low frequency AC field. Collective shuttling of entrained particles directly applies to the problem of diffusion of long molecular chains in zeolites and shape selective catalytic reactions in living cells. In particular, we stress that collective shuttles can be much more efficient than some other shuttle mechanisms, as they allow one to control the rate of transport by adding (removing) a single molecule to (from) the molecular chain.

The present paper is organized as follows: Our main goal is to analyze the effects of the pair attraction between the particles on the rectification current of single-files of colloidal particles which are confined within a narrow channel with corrugated walls and subjected to DC or AC drives. The general theoretical framework for our analysis of our model system, which is based on DDFT, is presented in Sec.\,\ref{ddft}. The one body density distribution $\rho(x,t)$ of the diffusing particles, which is a function of position $x$ and time $t$, obeys a nonlinear Fokker-Planck equation and DDFT \cite{marini99, marini00, archer04, archer04b} provides a closure approximation that allows us to solve for the dynamics of $\rho(x,t)$. The DDFT dynamical equation for the system takes the form of a conserved gradient dynamics, which requires as input a suitable approximation for the Helmholtz free energy functional for the system \cite{marini99, marini00, archer04, archer04b}. In the presence of a DC drive, the free energy contains a potential energy term proportional to $x$ that acts as a continuous external energy source, i.e., the system remains permanently out of thermodynamic equilibrium. In consequence, the free energy of a single-file of interacting particles is not necessarily a monotonically decreasing function of time. Therefore, the system can exhibit stable time-periodic density profiles and currents. In particular, in Secs\,\ref{cond} and \ref{DC} we discuss the relation of the onset of time-periodic density variations to the condensation of particles into compact clusters and the depinning of these clusters from the channel corrugations.

In Sec.\,\ref{zerolength} we examine the dynamics of point-like particles and we demonstrate that spontaneous symmetry breaking induced by attraction leads to the coexistence of stable time-periodic and stationary densities. Multistability of the long-time density distributions indicates that for the same combination of parameters in our model, the channel can operate in two different regimes, transporting the particles with either high or low efficiency. For finite-size particles the range of values of the system parameters that allows for time-periodic density profiles, is much broader than for point-like particles, as explained in Sec.\,\ref{finitelength}.

In Sec.\,\ref{acdrive} we discuss the low frequency rectification current for particles driven by an AC (square wave) drive through a channel with a spatially asymmetric potential. In Sec.\,\ref{currmax} we show that the effect of the spatial asymmetry is that the rectification current may be maximized, using the strength of the attraction between the particles as the control parameter. In Sec.\,\ref{strongatt} we derive an effective equation of motion for a condensate in the limit of infinitely strong attraction between the particles. We show that the low frequency transport efficiency can be increased by several orders of magnitude for strongly attracting particles as compared to non-interacting particles.
Finally, in Sec.\ \ref{conc} we close with a few concluding remarks.

\section{DDFT for interacting hard rods in a periodic external potential}
\label{ddft}
When finite sized colloids are confined within a long narrow channel, the particle motion becomes one dimensional, and the particles can be modeled as 1D hard rods of length $h$. We model the dynamics of $N$ hard rods using overdamped stochastic equations of motion. The particles move in a channel of total length $S$ and interact with the channel walls via a periodic corrugated potential $U(x)$ with spatial period $L$. In a system with periodic boundary conditions (e.g., a circular geometry), $S$ is an integer multiple of $L$, i.e. $S = M L$. The integer $M$ determines the average number of particles per unit cell of length $L$, to be $N/M$. To ensure that the combined length of the $N$ rods is smaller than the total system size, we require $N h \le S$.

The total instantaneous potential energy of the $N$ rods, which move in the periodic external potential $U(x)$ under the action of a time-dependent external drive $A(t)$, is:
\begin{eqnarray}
\label{potential}
\Phi(\{x_j\},t)=\sum_i[U(x_i)-A(t)x_i]+\frac{1}{2}\sum_{i,j \neq i} w(| x_i -x_j |),
\end{eqnarray}
where $w(x)$ denotes the interaction potential between a pair of particles $i$ and $j$, where $i,j=1,\dots,N$, which are located at positions $x_i$ and $x_j$, respectively, and are separated by the distance $x=|x_i -x_j|$. In general, the potential $w(x)$ can be decomposed into two terms. One accounts for the attraction (denoted by subscript ``at'') between the particles and the other for the hardcore repulsion (subscript ``hc''), i.e., $w(x) = w_{\rm at}(x) + w_{\rm hc}(x)$. The hard-core repulsive potential ensures that the rods are impenetrable, that is,
\begin{eqnarray}
\label{hr}
w_{\rm hc}(x) =
\left\{
\begin{array}{c}
\infty,\,\,\,x<h \\
0,\,\,\,\,\,x \geq h.
\end{array}
\right.
\end{eqnarray}
We assume that $w_{\rm at}(x)$ becomes negligible at distances $x$ much larger than a certain effective interaction range $l_{\rm int}$.

The overdamped dynamics of the particles is described by $N$ coupled Langevin equations
\begin{eqnarray}
\label{langevin}
\frac{1}{\Gamma}\frac{\partial x_i}{\partial t} = -\frac{\partial \Phi(\{x_j\},t)}{\partial x_i} + \sqrt{2T}\xi_i(t),
\end{eqnarray}
where $T$ is the temperature of the system and $\xi_i(t)$ are independent Gaussian white noises with the correlation functions $\langle \xi_i(t) \xi_j(t^\prime) \rangle = \delta_{ij}\delta(t-t^\prime)$. Henceforth we set the solvent friction constant $\Gamma=1$.

In the case of a circular geometry, Eqs.\,(\ref{langevin}) are supplemented with periodic boundary conditions (BC) with the period equal to the system size $S$. This requires that all functions in Eqs.\,(\ref{langevin}) are periodic with period $S$. This is not the case if the contribution to the interaction potential $w(x)$ from the potential $w_\mathrm{at}(x)$ is long ranged, as particles would interact then with themselves. However, $w(x)$ can be made compatible with the desired periodic BC by taking the interaction range $l_{\rm int}$ to be sufficiently small compared with the system size. Therefore, throughout we impose the condition $w_{\rm at}(S) \approx 0$.

The Fokker-Planck (Smoluchowski) equation for the time evolution of the one body density distribution $\rho(x,t)$ is \cite{marini99, marini00, archer04, archer04b}:
\begin{eqnarray}\notag
\frac{\partial \rho(x,t)}{\partial t} &=& T \frac{\partial^2 \rho(x,t)}{\partial x^2} 
+\,\frac{\partial}{\partial x}\Bigg[ \rho(x,t) \frac{\partial U_{\rm eff}(x,t)}{\partial x}\Bigg] \notag \\
&\,&+\frac{\partial}{\partial x}\Bigg[ \int_{-\frac{S}{2}}^{\frac{S}{2}} {\rm d} x' \rho^{(2)}(x,x',t) \frac{\partial}{\partial x} w(|x-x'|)\Bigg],
\label{eq:FP}
\end{eqnarray}
where $\rho^{(2)}(x,x',t)$ is the non-equilibrium two-body distribution function for the particles in the system and $U_{\rm eff}(x,t)=U(x)-A(t) x$ is the effective external potential. Note that the density profile $\rho(x,t)$ is normalized, so that the spatial integral over $\rho(x,t)$ is equal to the total number of particles in the system; i.e.\ $\int_{-S/2}^{S/2}\rho(x,t)\,dx = N$.

In order to solve Eq.\ \eqref{eq:FP}, a suitable closure approximation for $\rho^{(2)}(x,x',t)$ is required. The approach taken in DDFT is to approximate $\rho^{(2)}(x,x',t)$ by the two-body distribution function of an equilibrium fluid with the same one-body density profile as the non-equilibrium system \cite{marini99, marini00, archer04, archer04b}. This closure relates the integral in the final term in Eq.\ \eqref{eq:FP} to the functional derivative of the excess part of the Helmholtz free energy functional $F[\rho]$, which is the central quantity of interest in equilibrium density functional theory \cite{hansen2006tsl, evans1992fif, evans}. Making this approximation yields the following equation for the dynamics of the one-particle density distribution $\rho(x,t)$:
\begin{eqnarray}
\label{eq3}
\frac{\partial \rho(x,t)}{\partial t} = \frac{\partial}{\partial x}\left[ \rho(x,t)\frac{\partial }{\partial x}\frac{\delta F[\rho(x,t)]}{\delta \rho(x,t)}\right].
\end{eqnarray}
For the case with periodic BC, the Helmholtz free energy functional $F[\rho]$ is of the form \cite{evans1992fif, hansen2006tsl}:
\begin{eqnarray}
\label{dft_energy}
F[\rho(x,t)] &=& T \int_{-\frac{S}{2}}^{\frac{S}{2}} dx\,\rho(x,t)[\ln{\rho(x,t)}-1] \nonumber \\
&+& \int_{-\frac{S}{2}}^{\frac{S}{2}} dx\,U_{\rm eff}(x,t)\rho(x,t) \nonumber \\
&+&  F_{\rm hc}[\rho] + F_{\rm at}[\rho],
\end{eqnarray}
where the first term on the right hand side is the ideal-gas contribution to the free energy, $F_{\rm hc}$ is the excess contribution to the free energy due to the hardcore repulsion between the particles, and $F_{\rm at}$ represents the contribution due to the attractions between the particles. In a mean-field approximation \cite{evans1992fif}, $F_{\rm at}$ is given by
\begin{equation}
\label{eq:F_at}
F_{\rm at}[\rho] = \frac{1}{2}\int_{-\frac{S}{2}}^{\frac{S}{2}} dx \int_{x-\frac{S}{2}}^{x+\frac{S}{2}} dx'\,w_{\rm at}(\mid x-x' \mid)\rho(x)\rho(x').
\end{equation}
The exact expression for the equilibrium excess Helmholtz free energy for hard rods of length $h$, $F_{\rm hc}[\rho]$, was first presented in Ref.\,\cite{percus76}. The result is:
\begin{eqnarray}
\label{hr_energy}
F_{\rm hc}[\rho] = \frac{1}{2}\int_{-\frac{S}{2}}^{\frac{S}{2}} dx\, \phi[\rho(x)] \left\{\rho\left(x+\frac{h}{2}\right)+\rho\left(x-\frac{h}{2}\right)\right\},
\end{eqnarray}
where $\phi[\rho(x,t)] = -T\ln{[1-\eta(x,t)]}$ and $\eta(x,t) = \int_{x-h/2}^{x+h/2}dx'\,\rho(x',t)$. It should be emphasized that the functional in Eq.\,(\ref{hr_energy}), is strictly only exact for 1D equilibrium systems of hard-rods treated in the grand canonical ensemble \cite{percus76, evans1992fif}. However, as the (average) number of particles in a system is increased, the difference between results from treating a system canonically or grand canonically diminishes, so that the theory can safely be extended to describe single-files with a fixed but large numbers of rods.

In earlier work, the DDFT approach was used to study the dynamics of an ensemble of pure hard rods (i.e.\ with no attractive interactions) \cite{marini99, marini00, penna03}. More recently \cite{PA10}, we have applied the DDFT formalism to describe a file of hard rods interacting via a pair potential with an attractive contribution. In both cases, the free energy functional given in Eq.\,(\ref{hr_energy}) was shown to reproduce fairly closely the results from Brownian dynamics computer simulations [i.e.\ results from numerically integrating Eqs.\ \eqref{langevin}], even for relatively small numbers of particles, $N \gtrsim 10$.

Using Eqs.\,(\ref{dft_energy}) and (\ref{hr_energy}), Eq.\,(\ref{eq3}) can be rewritten in the form of a conservation law, i.e. in terms of the instantaneous current density $J(x,t)$,
\begin{eqnarray}
\label{eq:continuity}
\frac{\partial \rho(x,t)}{\partial t} &=& -\frac{\partial J(x,t)}{\partial x},
\end{eqnarray}
where
\begin{eqnarray}
\label{current}
J(x,t) &=&   \rho(x,t)\left[ -T \frac{\partial }{\partial x}\ln \rho(x,t) -\frac{dU(x)}{dx}+A(t) \right.\nonumber \\
&-&  T\left( \frac{\rho(x+h,t)}{1-\eta(x+h/2,t)} - \frac{\rho(x-h,t)}{1-\eta(x-h/2,t)}\right)  \nonumber \\
&-&  \left. \int_{x-\frac{S}{2}}^{x+\frac{S}{2}}dx^\prime\,\rho(x^\prime,t)\frac{\partial w_{\rm at}}{\partial x}(x-x^\prime) \right].
\end{eqnarray}
Before we discuss our results for solutions of Eqs.~(\ref{eq:continuity}) and (\ref{current}), we would like to point out that we see many parallels between the dynamics described by the partial integro-differential Eqs.~(\ref{eq3}) with (\ref{dft_energy}) and the dynamics of partially wetting drops and films on solid substrates described by so-called thin film equations (fourth order partial differential equations) \cite{KaTh07}. Formal similarities between thin film equations and some Fokker-Planck equations for interacting particles have recently been pointed out in the case of spatially-asymmetric ratchets with a temporal AC drive \cite{ThJo10}. In the present case, we find that some of our results are similar to results for the case of liquid drops on horizontal \cite{TBBB03} and inclined \cite{dep2} heterogenous solid substrates. These similarities arise due to (i) the similar gradient dynamics form of the evolution equation for the conserved field (here $\rho$), (ii) the presence of similar physical effects. The role of the attractive and repulsive force between particles is taken by the partial wettability of the liquid \cite{deGe85}; the stabilizing role of diffusion is played by surface tension; the channel corrugations are similar to substrate heterogeneities; the DC drive corresponds to constant driving parallel to the substrate (e.g., drop on an incline); and the present AC drive is similar to substrate vibrations \cite{JoTh10} or oscillating electric fields \cite{JoTh07}.

Here, we solve Eq.\,(\ref{eq:continuity}) imposing periodic BC on the domain $x\in [-S/2,S/2]$ which is centered at the origin, $x=0$. Due to the external driving force the system remains permanently out of thermodynamic equilibrium. Therefore, in an infinite domain or in a finite domain with periodic BC, the non-equilibrium dynamics of the system with DC drive is not relaxational, in contrast to the case of zero-flux BC that would correspond to a closed finite system \cite{marini99}. Note, however, that the channel corrugations may result in local equilibria. An AC drive keeps the system out of equilibrium for any BC.  The non-relaxational character is reflected by the finding that with periodic BC, the free energy $F[\rho(x,t)]$ in Eq.~(\ref{dft_energy}) is not necessarily a monotonically decreasing function of time, i.e., it does not play the role of a Lyapunov functional for the gradient dynamics Eq.~(\ref{eq3}). This can best be seen by considering the total time derivative of the free energy
%
\begin{eqnarray}
\label{dfdt}
\frac{d F[\rho]}{d t} &=& \int_{-\frac{S}{2}}^{\frac{S}{2}}\frac{\delta F}{\delta \rho }\frac{\partial \rho(x,t)}{\partial t}\,dx \nonumber \\
&=& \int_{-\frac{S}{2}}^{\frac{S}{2}}\frac{\delta F}{\delta \rho }\frac{\partial}{\partial x}\left[\rho(x,t)\frac{\partial }{\partial x}\frac{\delta F}{\delta \rho}\right] \,dx \nonumber \\
&=&\left[\frac{\delta F}{\delta \rho}\rho \frac{\partial }{\partial x} \frac{\delta F}{\delta \rho}\right]_{-\frac{S}{2}}^{\frac{S}{2}}- \int_{-\frac{S}{2}}^{\frac{S}{2}}\rho \left[\frac{\partial }{\partial x}\frac{\delta F}{\delta \rho}\right]^2\,dx.
\end{eqnarray}
where we have used Eq.\,(\ref{eq3}) and integrated by parts. The functional derivative $\delta F/\delta \rho$ is not periodic in $x$, due to the ``tilted" effective potential $U_{\rm eff} = U(x)-A(t)x$, and so the boundary term in the last line of Eq.\,(\ref{dfdt}) does not normally vanish. It is equal to $AS J_{S/2}>0$, where $J_{S/2}$ is the current density on the boundary, and $J_{S/2}<0$ for $A<0$.
It then follows that for $A \not= 0$ the time derivative  $d F[\rho(x,t)]/d t$ is not necessarily a negative quantity. This allows for time-oscillatory behavior of the system, even in the long-time limit. In other words, Eq.\,(\ref{eq:continuity}), with $A(t)=A$, may admit stable cyclo-stationary or time-periodic solutions, the existence of which will be discussed in the next section.

For stable time-periodic solutions, the average particle current $\bar{J}$ is obtained from Eq.\,(\ref{current}) by averaging $J(x,t)$ over position and over time, namely,
\begin{eqnarray}
\label{eq8}
\bar{J} = \frac{1}{\tau}\int_{-\frac{S}{2}}^{\frac{S}{2}}dx\,\int_{t^\prime}^{t^\prime + \tau}dt\,J(x,t),
\end{eqnarray}
where $\tau$ is the period of the oscillations. Throughout this paper we will use the average current per particle $J$, which is related to the total current in Eq.\,(\ref{eq8}) via $J = \bar{J}/N$.

\section{Spontaneous condensation}
\label{cond}
In the absence of the channel potential and without AC or DC drive, i.e for $U(x)_{\rm eff}=0$, the DDFT equation (\ref{eq3}) admits a stationary homogeneous solution $\rho(x)=\rho_0$, with constant average density $\rho_0 = N/S$. Because of the competition between the destabilizing attractive forces and the stabilizing effect of the thermal motion of the particles (diffusion), the homogeneous density distribution is not always stable. In order to minimize their free energy, the attracting particles tend to condense into a set of clusters. This tendency is opposed by the stabilizing action of diffusion, which leads to a spreading of the particles away from one another. Depending on which process dominates, the homogeneous state $\rho(x)=\rho_0$ may either be (linearly) stable or unstable. Note that in this case the dynamics is relaxational and the functional (\ref{dft_energy}) is a Lyapunov functional [cf.~Eq.~(\ref{dfdt})].

One may alternatively take a thermodynamic rather than a dynamical point of view for understanding this instability: Recall that the Helmholtz free energy of the system $F={\cal U}-T{\cal S}$, where ${\cal U}=\langle\Phi\rangle$ is the internal energy, ${\cal S}$ is the entropy of the system and $\langle \cdots \rangle$ denotes a statistical average \cite{hansen2006tsl}. 
At equilibrium, $F$ is minimal. At high temperatures $T$, this is achieved by maximizing ${\cal S}$ (i.e.\ by dispersing the particles throughout the system in a maximally disordered way), since the term $-T{\cal S}$ is the dominant contribution to the free energy at high temperatures $T$. However, at low temperatures, the internal energy contribution ${\cal U}$ dominates the free energy $F$ and so the attracting particles minimize the free energy by gathering together to minimize the internal energy ${\cal U}$. 

We model the attractive contribution to the pair potential between the particles by a simple exponential function of the form
\begin{eqnarray}
\label{attractive_potential}
w_{\rm at}(x) =-\alpha \exp{(-\lambda x)},
\end{eqnarray}
where the parameters $\alpha$ and $\lambda$ characterize the strength of attraction and the attraction range $l_{\rm at} = 1/\lambda$, respectively. It should be noted that the mean field approximation for the contribution to the Helmholtz free energy due to the attractive interactions \eqref{eq:F_at} used here, is only quantitatively reliable when $\lambda h \lesssim 1$, i.e.\ when any given particle is interacting with several of its neighbors so that a mean-field approximation is appropriate. However, even when the attraction range is somewhat shorter than this, the mean-field approximation remains qualitatively correct.

In order to study the stability of the homogeneous state, we linearize Eq.\,(\ref{eq3}) using the standard plane wave anstatz $\rho(x,t) = \rho_0 + \varepsilon e^{\beta(k)t + ikx}$, where the sign of the growth rate $\beta(k)$ determines the stability of the homogeneous solution $\rho_0$. Note that for the stable fluid the dispersion relation $\beta(k)$ is closely related to the static structure factor $S(k)$, via $\beta(k)=-T k^2/S(k)$ \cite{archer04}. Neglecting exponentially small terms of the order of $\exp{(-\lambda S)}$, we obtain to leading order in $\varepsilon$
\begin{eqnarray}
\label{disp}
\beta(k) &=& -Tk^2 - \frac{2Tk\rho_0\sin{(kh)}}{1-\rho_0 h} -\frac{4 T\rho_0^2[\sin{(kh/2)}]^2}{(1-\rho_0 h)^2} \nonumber \\
&+& \frac{2\alpha \lambda \rho_0 k^2}{k^2+\lambda^2}.
\end{eqnarray}
\begin{figure}[t]
\centering
\includegraphics[width=0.48\textwidth]{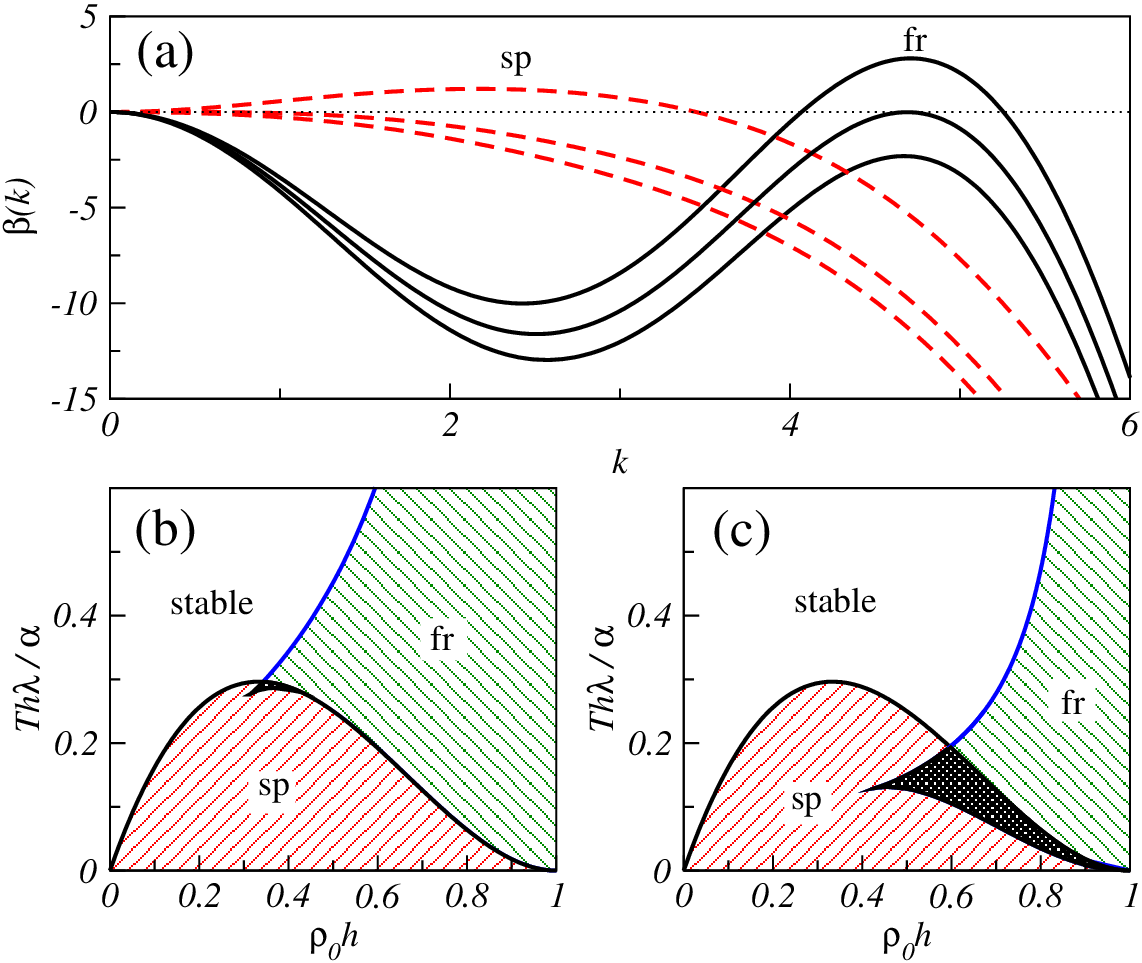}
\caption{(Color online)  (a) Typical dispersion relations $\beta(k)$ close to the onset of the spinodal mode (sp), as shown by dashed lines and the freezing mode (fr), as shown by solid lines. The $\beta(k)$ curves near the spinodal instability are for $\rho_0=0.2$ and $\alpha=10$, $11.8$ and $17$, which are close to the onset of the instability. The curves for the freezing instability are for $\rho_0=0.6$ and $\alpha=4$, $4.9$ and $6$.  The other parameters are $T=1$, $\lambda=3$, $h=1$. (b-c) Stability diagram for the system with uniform constant density $\rho(x)=\rho_0$, in the plane spanned by $\widetilde{T}=Th\lambda/\alpha$ and $\tilde\rho=\rho_0 h$ for (b) $\xi=\lambda h=3$ and (c) $\xi=1.5$. The labels ``sp'' and ``fr'' denote (hatched) regions where the uniform density is only linearly unstable to the spinodal and freezing instability, respectively. In the cross-hatched region both modes are unstable.\label{F0}}
\end{figure}

Inspection of Eq.\,(\ref{disp}) shows that two different scenarios exist where the homogeneous density $\rho_0$ is linearly unstable. On the one hand, the solution $\rho(x,t)=\rho_0$ can become unstable via the standard spinodal phase separation mechanism where the material separates into regions of low density (gas) and hight density (liquid), which we call the ``spinodal mode'' (denoted by ``sp''). It is associated with an instability against harmonic perturbations with wave numbers $0<k<k_0$, where $k_0$ is an upper value obtained by solving the equation $\beta(k_0)=0$. The spinodal instability sets in when the leading coefficient of the expansion of $\beta(k)$ in powers of $k^2$ (i.e., the coefficient of the $k^2$ term) vanishes, i.e., the wave number at onset is zero. This corresponds to a type $II_s$ instability in the classification of Cross and Hohenberg \cite{CrHo93}.  Based on the Taylor expansion of the relation (\ref{disp}) one obtains $\widetilde T=2\tilde\rho\,(1-\tilde\rho^2)$ as the condition for the onset of the spinodal mode [cf.~Fig.~\ref{F0}(b)]. Here, we have introduced the reduced temperature $\widetilde{T}=Th\lambda/\alpha$ and the reduced density $\tilde{\rho}=\rho_0 h$.  The critical point for decomposition is found at $\widetilde{T}_c= 8/27$ and $\tilde{\rho}_c=1/3$.

The stability condition is equivalent to the thermodynamic stability criterion requiring that the isothermal compressibility be negative. This corresponds to the Helmholtz free energy per unit length of the system becoming concave, namely, the boundary of the spinodal instability is given by $\delta^2 F[\rho] / \delta \rho^2|_{\rho_0} =0$. When $\delta^2 F[\rho] / \delta \rho^2|_{\rho_0} >0$ the uniform solution is linearly stable and when $\delta^2 F[\rho] / \delta \rho^2|_{\rho_0} <0$ the uniform solution is linearly unstable. The dispersion relation $\beta(k)$ close to the onset of the spinodal mode is shown in Fig.\ref{F0}(a) by dashed lines, which correspond to three different values of the attraction strength $\alpha$, chosen close to the spinodal instability threshold as obtained from Eq.\,(\ref{disp}). 

On the other hand, the solution $\rho(x,t)=\rho_0$ can become unstable via a freezing mode of the system, where the particles become localized and the density profile exhibits a series of sharp peaks separated by distances smaller than $L$. This instability corresponds to type $I_s$ in the classification of Ref.~\cite{CrHo93}. In the context of reaction-diffusion systems it is sometimes referred to as a Turing instability \cite{Nico95}.  This mode (which we denote by ``fr'') sets in at a non-zero critical wave number $k_c$, as illustrated in Fig.\ref{F0}(a) by the solid lines for $\rho_0 =0.6$. Beyond onset, the freezing mode gives rise to the growth of periodic modulations in the density profile with a wavelength $ \approx 2\pi/k_c$.



In Figs.\,\ref{F0}(b) and (c) we plot the linear stability diagram of the system with uniform density $\rho_0$, in the plane spanned by $\widetilde T =Th\lambda/\alpha$ and $\tilde\rho=\rho_0 h$ for the interaction length ratio (b) $\xi\equiv\lambda h=3$ and (c) $\xi=1.5$. The labels ``sp'' and ``fr'' mark regions where the uniform system is linearly unstable to the spinodal and freezing mode, respectively. As discussed above, the onset of the spinodal mode only depends on the reduced temperature $\widetilde T$ and  reduced density $\tilde\rho$. In contrast, the onset of the freezing mode also depends on the value of $\xi$.  For relatively long rods with $\xi=3$, where the attraction range is short compared to the core size $h$, the region of the freezing instability extends down to moderate values of the reduced density $\tilde\rho\approx 0.5$. For a reduced temperature above $\widetilde T_c=8/27$ only the freezing instability exists at large and moderate $\tilde\rho$. For small values of $\xi$, corresponding to the attraction range $\lambda^{-1}$ being large compared to $h$,  and for which the mean-field approximation for the free energy used is expected to be most reliable, the freezing mode is only found at extremely high packing fractions, $\tilde\rho \approx 1$.  At such high densities, the critical wave number $k_c$ of the freezing mode is approximately $k_c \approx 2\pi/h$, giving rise to the formation of density peaks separated by a distance $\approx h$, as one would expect for a frozen system.

Below $\widetilde T_c$, the spinodal mode exists for a range of $\tilde\rho$ that with decreasing $\widetilde T$ extends on both sides of the critical value $\tilde\rho =1/3$. This implies that the spinodal mode sets in at smaller and smaller values of the density as $\widetilde T$ is decreased. At large $\tilde\rho$ there exists a region where both linear modes are unstable [cross-hatched in Figs.\,\ref{F0}(b) and (c)]. The region where only the freezing mode exists is shifted towards higher densities as $\widetilde T$ decreases. The definition of $\widetilde T$  implies that for any given physical temperature $T$, a decrease in the interaction strength $\alpha$ below the threshold value $\alpha_c=27Th\lambda/8$ stabilizes the spinodal mode. 
For any finite temperature the system can be quenched into the freezing unstable region by increasing the average density in the system.

\section{DC drive}
\label{DC}
Having considered the stability of the uniform system, we now consider the non-uniform system that is subject to a periodic external potential $U(x)$ and the DC driving force $A$. We model the periodic potential, induced by the corrugated channel walls, by the standard bi-harmonic ratchet potential $U(x)=\sin{(2\pi x)}+0.25\sin{(4\pi x)}$ \cite{RMP09}. Recall that to drag a single particle over one of the barriers in $U(x)$, one must apply a force $A_R=3\pi$ to pull the particle over the barrier to the right and a force $A_L=3\pi/2$ to pull it to the left. $A_{L}$ and $A_R$ are termed the left and the right depinning thresholds, respectively \cite{RMP09}.  Note that all the results reported in this section for a DC drive remain qualitatively valid even for a simpler symmetric periodic external potential, such as $U(x)=\sin{(2\pi x)}$. The case of a periodic external potential without and with DC drive shows some similarities to liquid drops/films on periodically heterogeneous substrates without \cite{TBBB03} and with \cite{dep1} a driving force parallel to the substrate, respectively. The former case will be explored elsewhere. Below we discuss similarities and differences for the case with DC driving.

\subsection{Zero rod length and finite interaction range}
\label{zerolength}
As a reference system we first consider a file of point-like particles, i.e., with $h=0$, interacting solely via the exponential soft core potential $w_{\rm at}(x)$ and driven by a DC external force. When the characteristic interaction range $l_{\rm at}$ between the point-like rods is small, i.e.\ when $l_{\rm at}=1/\lambda \rightarrow 0$, a local approximation can be made for the dynamical equations for the system, as shown in Refs.\,\cite{sav04,sav05}. In this limit, the integral involving $w_{\rm at}$ in Eq.\,(\ref{current}) can be reduced to a local function of the form $\sim g\rho(x,t)\partial \rho(x,t)/\partial x$, where the coefficient $g$ is a parameter determined by the strength of the interactions between the particles. In Refs.\,\cite{sav04,sav05} it was shown that when $g$ is increased beyond a certain critical value, the density distribution of the attracting particles exhibits a spontaneous symmetry breaking transition, where the stationary periodic density profile with period $L$ [the period of the modulations in $U(x)$] becomes unstable and evolves toward a stable stationary distribution with period $S$ (the total system length). We now go beyond the analysis of Refs.\,\cite{sav04,sav05} and consider such a symmetry breaking mechanism for particles interacting via a potential with a nonzero interaction range, $l_{\rm at} \not= 0$.


For convenience we fix the average particle density to be $\bar \rho L = 1$, corresponding to one particle per period $L$ of the channel, and we set the constant drive, $A=-1$. Using the numerical continuation package {\it AUTO} \cite{AUTO}, we follow the branch of solutions corresponding to a stationary density distribution $\rho_s(x)$, that originates from the stationary density profile for the case when $\alpha=0$ (i.e.\ a non-interacting ideal-gas of particles). Note that in the long-time limit, the density profile for the ideal-gas remains stable and stationary, regardless of the form of the channel potential, $U(x)$, the magnitude of the drive, $A$, or the temperature of the system, $T$ \cite{risken,reimann02}.

We determine the stationary solutions of Eq.\ (\ref{eq:continuity}) with the current given by Eq.\,(\ref{current}) that contains nonlocal terms, by means of the Fourier mode method described in Ref.\,\cite{bordyugov07}. The density profile is discretized over the domain $[-S/2,S/2]$, derivatives are obtained using finite difference approximations, and the non-local terms are calculated using a Fast Fourier transform. We start from the equation for the stationary solution of Eq.\,(\ref{eq:continuity}), $\partial J(x,t)/\partial x =0$, which is then written as a set of algebraic equations for the Fourier components of the current $J(x,t)$ and from this we obtain our solutions for the stationary density profile $\rho_s(x)$. Using the continuation package {\it AUTO} allows us to detect the presence of Hopf bifurcations as well as to trace the solution branches for both the stationary solutions and the time-periodic ones that emerge from them.

\begin{figure*}[t]
\centering
\includegraphics[width=0.98\textwidth]{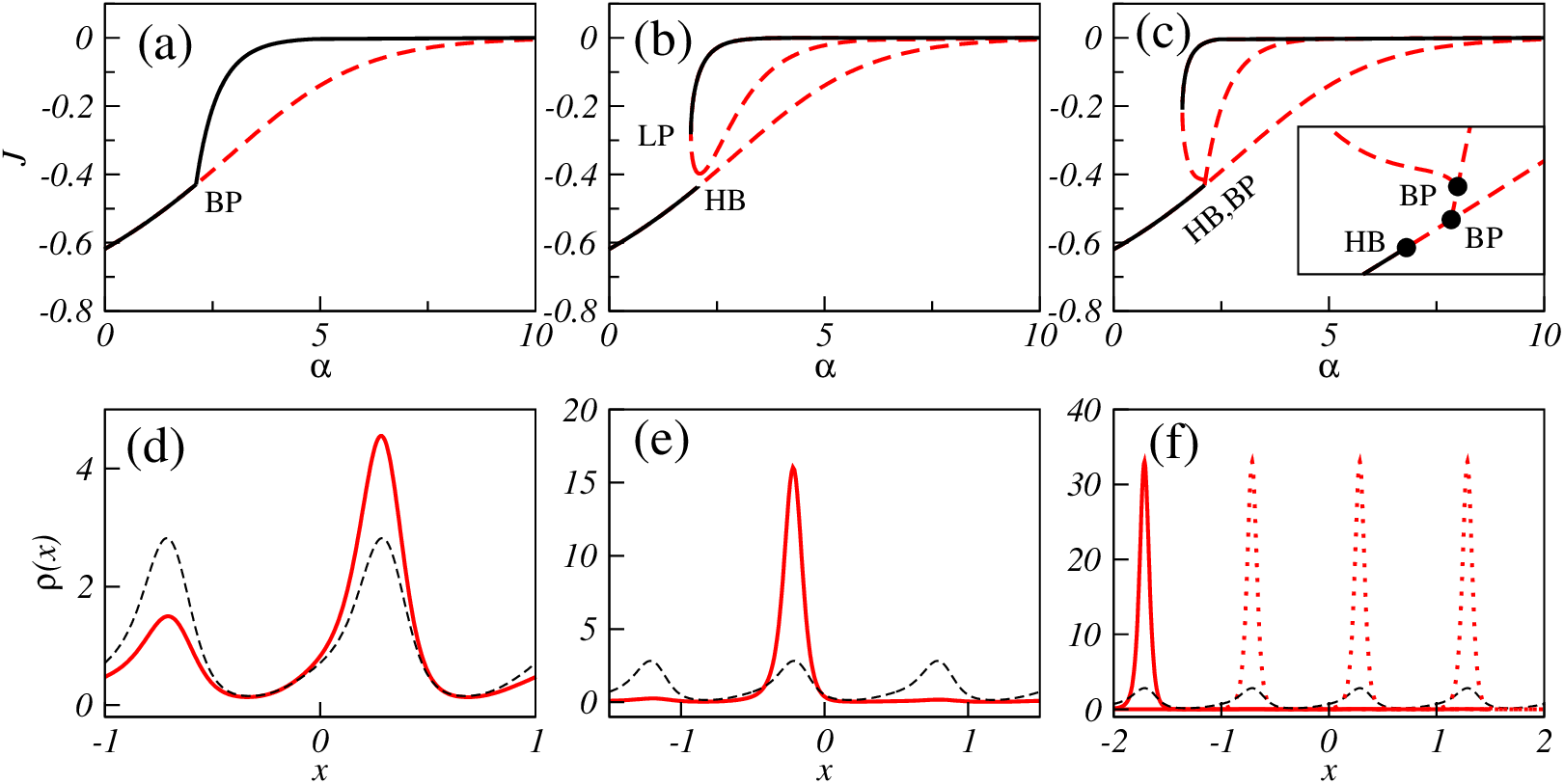}
\caption{(Color online)  Panels (a), (b) and (c) display bifurcation diagrams for the stationary density distributions in terms of the average current $J$ versus the interaction strength $\alpha$, for $\lambda=5$, $\bar \rho=1$, $A=-1$, $h=0$ and total system length (a) $S=2L$, (b) $3L$ and (c) $4L$. The solid and dashed lines correspond to stable and unstable solutions, respectively. The points labeled ``HB'' and ``BP'' denote Hopf bifurcation and branching points, respectively. Panels (d), (e) and (f) display selected corresponding density profiles. The solid lines are profiles with spatial periods $2L$, $3L$ and $4L$, respectively, for $\alpha=2.2$. The dashed lines represent the unstable solution with period $L$. In (f) the solid and dotted lines represent all four possible $4L$-periodic solutions. \label{F1}}
\end{figure*}

We begin by discussing the bifurcation diagrams of the stationary solutions of Eq.\,(\ref{eq:continuity}) on varying the interaction strength, $\alpha$, for a fixed value of the range parameter of the pair potential, $\lambda=5$, and the fixed $U(x)$. These are shown in Figs.\,\ref{F1}(a)-(c), for three different systems with lengths, $S=2L$, $3L$ and $4L$, respectively. The solid lines correspond to stable solutions and the dashed lines to unstable (saddle point) solutions. The labels ``HB'' and ``BP'' stand for Hopf bifurcation and branching point, respectively.

As the interaction strength is increased beyond a critical value $\alpha_c$, the $1L$-periodic solution that is stable for small $\alpha$ becomes unstable either via a (period-doubling) pitchfork bifurcation (for $S=2L$), or via a Hopf bifurcation (for $S>2L$). In the case of the pitchfork bifurcation, displayed in Fig.\,\ref{F1}(a), a double branch of stable solutions emerges at the bifurcation point. The two branches are related by the discrete translation symmetry $x\to x+L$ and can therefore not be distinguished in Fig.~\ref{F1}(a).  This new branch corresponds to a solution with a larger spatial period, equal to the system size $S=2L$, and a smaller value for the particle current $J$. One may say that for $\alpha>\alpha_c$, the periodic potential is not strong enough to pin the clusters against their natural tendency to coarsen. For $S=2L$, we know on general grounds that [in a homogeneous system without driving $A=U(x)=0$] there are two possible coarsening modes: a translation mode, where the two clusters move towards each other, and a volume transfer mode, where material is transfered from one cluster to the other \cite{KaTh07}. It is known that both are stabilized by substrate heterogeneities
exerting a strong enough pinning influence \cite{TBBB03}. Not much is known, however, for driven systems ($A\neq0$). Our DDFT simulations show that in the present system, the instability is related to the volume transfer mode of coarsening.  The corresponding stable (solid line) and unstable (dashed line) stationary density profiles are displayed in Fig.\,\ref{F1}(d) for the case when $\alpha=2.2$.

The bifurcation diagram for the system with length $S=3L$ is qualitatively different from the one for the $S=2L$ case, as can be seen in Fig.\,\ref{F1}(b). One observes that the $1L$-periodic solution becomes unstable via a Hopf bifurcation. There exist branches of stationary solutions where the $x\to x+L$ translational symmetry is broken. However, they do not touch the primary branch of the $1L$-periodic solutions but are generated through a saddle-node bifurcation at the point marked by ``LP''. Solutions on these branches have period $S=3L$ and are either stable (upper branch) or unstable (lower branch). An example of a stable $3L$-periodic solution for $\alpha=2.2$ is displayed in Fig.\,\ref{F1}(e).

The bifurcation scenario for $S=4L$, displayed in Fig.\,\ref{F1}(c), is substantially more complex. The $1L$-periodic solution becomes unstable through a Hopf bifurcation. Very close to the HB point, the unstable stationary solution undergoes a primary period-doubling pitchfork bifurcation BP. Note that this first BP point for $S=4L$ coincides as expected with the BP point for $S=2L$. The newly formed $2L$-periodic solution is unstable and undergoes a further period-doubling pitchfork bifurcation at a second BP, which lies very close to the first BP, as shown in the inset of Fig.\,\ref{F1}(c).

For interaction strengths significantly larger than the critical values corresponding to the BP and the HB points, the only stable stationary solution of the DDFT equation (\ref{eq3}) has a period equal to the total system size $S$. Note that the multiplicity of the branch of the solutions with broken $x\to x+L$ symmetry depends on the total system size. For instance, for $S=4L$ there exist four such branches with spatial period $S$, as can be seen in Fig.\,\ref{F1}(f). Each solution exhibits a prominent maximum centered around one of the four minima of the external potential $U(x)$.  The solutions on the 4 branches are related by the symmetry $x\to x+L$. Therefore all of them correspond to the same value for the particle current $J$ and they can not be distinguished from one another in  Fig.\,\ref{F1}(c). 

From the results displayed in Fig.\,\ref{F1} we may draw two important conclusions: First, the detected Hopf bifurcations of the pinned stationary solutions signals the onset of time-periodic solutions of the DDFT equation (\ref{eq3}), even in the presence of a time independent drive. Second, for certain values of the interaction strength $\alpha$, two stable stationary solutions may coexist, giving rise to current multiplicity.
\begin{figure}[t]
\centering
\includegraphics[width=0.48\textwidth]{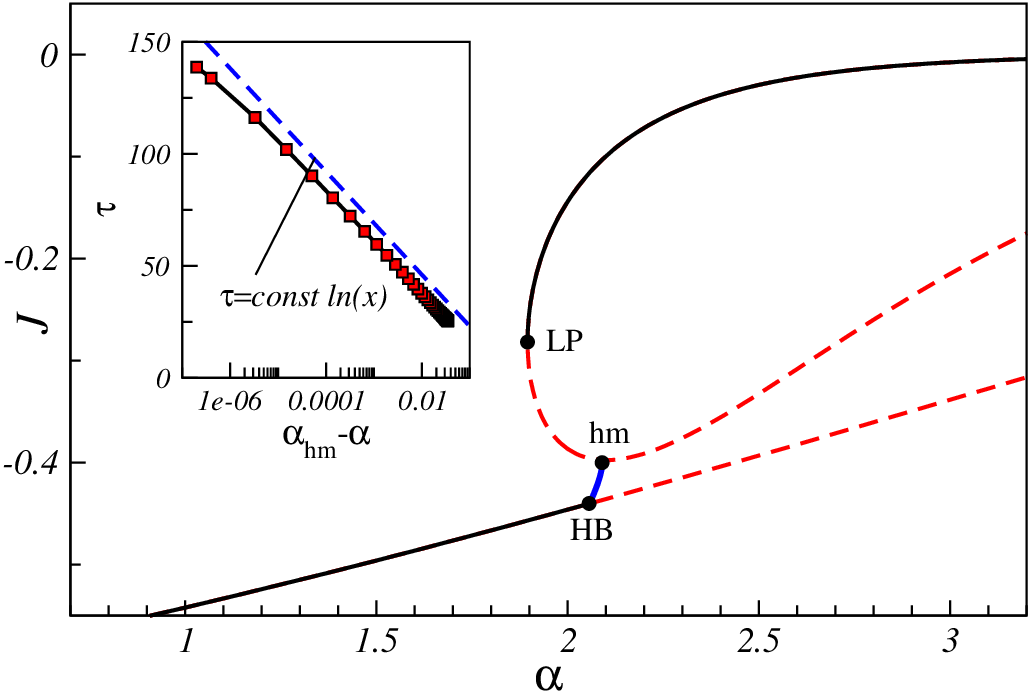}
\caption{(Color online)  Magnification of the region of Fig.\,\ref{F1}(b) close to HB point. The line connecting HB and ``hm'' corresponds to stable time-periodic solutions of Eq.\,(\ref{eq3}). The label ``hm'' stands for homoclinic bifurcation point. (Inset) Temporal period $\tau$ of the time-periodic solutions as a function of $\alpha_\mathrm{hm}-\alpha$. \label{F2}}
\end{figure}

These two findings are illustrated in detail in Fig.\,\ref{F2}, where we display a magnification of the region close to the bifurcations in Fig.\,\ref{F1}(b). In addition to the Hopf bifurcation (HB) and the saddle-node bifurcation (LP) of the stationary $3L$-periodic solutions, we display the branch of time-periodic solutions of Eq.\,(\ref{eq3}). It emerges at the HB point and terminates in a homoclinic bifurcation (labeled by ``hm'') where the time-periodic solution (limit cycle) collides simultaneously with all three unstable $3L$-periodic solution (unstable equilibria) \cite{Stro94}. The inset of Fig.\,\ref{F2} gives the temporal period $\tau$ as a function of the distance to the homoclinic bifurcation $\alpha_\mathrm{hm}-\alpha$. It shows a logarithmic dependence as expected close to a homoclinic bifurcation. We emphasize that these time-periodic solutions are stable, i.e., the corresponding Floquet multipliers are always located within the unit circle (not shown). Note that for clarity we not only suppress the branch of time-periodic solutions in Fig.\ \ref{F1}(b) but also a similar branch in Fig.\ \ref{F1}(c), for the system with $S=4L$.

In nonequilibrium driven systems, the loss of stability of the stationary solutions and the appearance of time-periodic solutions with a larger mean flow is sometimes associated with the concept of depinning. For example, in the study of liquid droplets on an inclined heterogeneous solid substrate, the dynamics of drop depinning has been studied in great detail -- see for example Refs.\ \cite{dep1,BKHT11} and references therein. In this situation the depinning is generally a transition from a steady droplet, pinned by the heterogeneity of the substrate, to a moving droplet, sliding down the incline under the action of gravity (or other driving forces parallel to the substrate). The depinning is usually investigated by increasing the driving force with all other parameters kept fixed. In such a case the dominant depinning mechanism is often related to a Saddle Node Infinite PERiod (sniper) bifurcation, although depinning via a Hopf bifurcation may also be observed in certain parameter regions \cite{dep1,BKHT11,dep2}. The depinning exhibited by the present system is observed when increasing the particle attraction $\alpha$, for a fixed value of the external drive $A$ and potential $U(x)$. This would correspond to a decrease in wettability for a droplet depinning in the thin film model.
Note also that this collective depinning is very distinct from the $T=0$ transition that is also referred to as `depinning', when the drive on a single particle exceeds either $A_L$ or $A_R$, the left and right single particle depinning thresholds \cite{RMP09}.

\begin{figure}[t]
\centering
\includegraphics[width=0.48\textwidth]{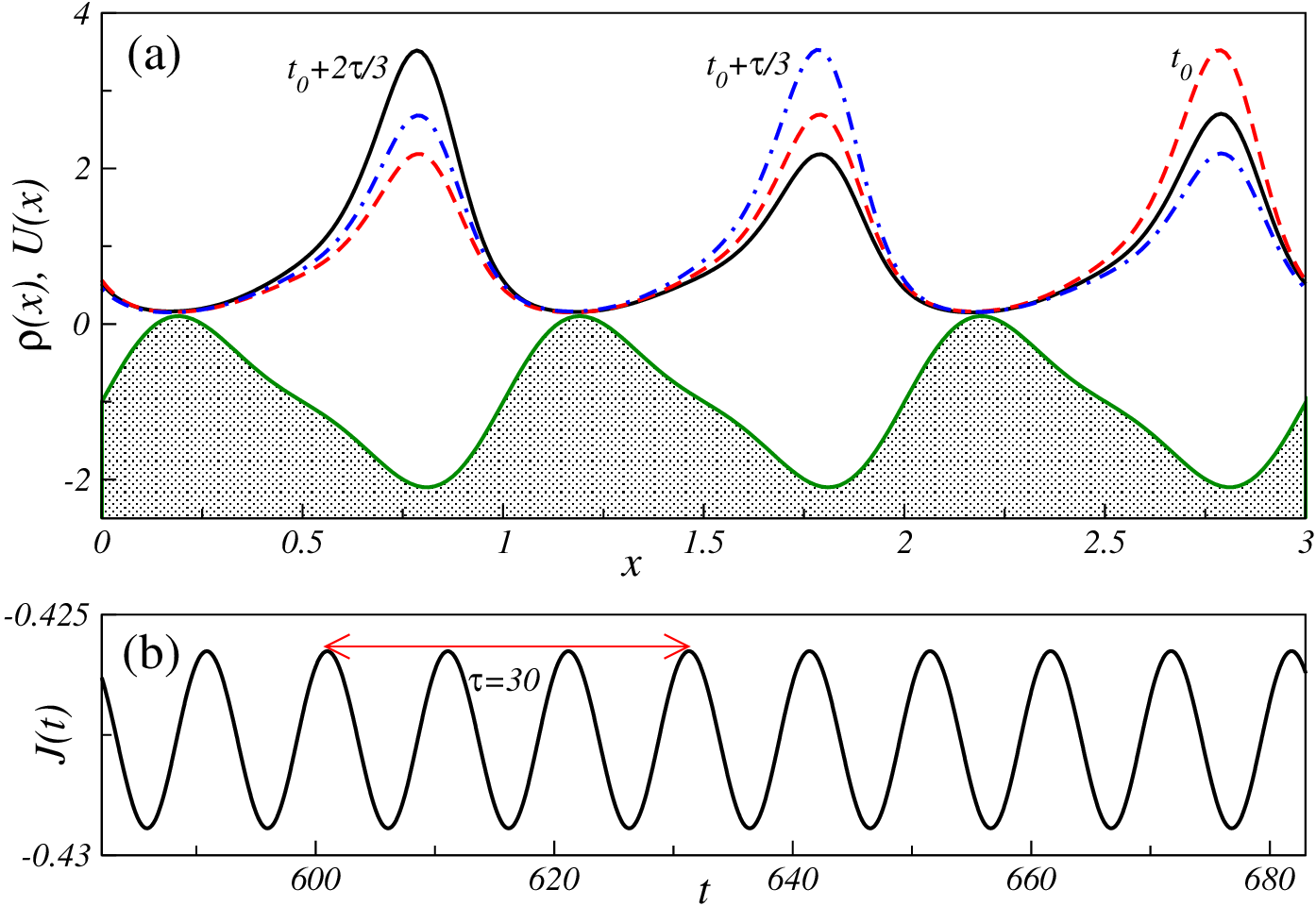}
\caption{(Color online) (a) Snapshots of the time-periodic density, $\rho(x,t)$, for $\alpha=2.08$ at three times, $t_0$ (dashed line), $t_0+\tau/3$ (dot-dashed line), and $t_0+2\tau/3$ (solid line). The temporal period of the solution is $\tau=30$; the remaining numerical parameters are as in Fig.\,\ref{F1}(b). The channel potential $U(x)$ is also displayed (heavy solid line limiting the shaded area). (b) The time dependent current $J(t) = (1/N)\int_{-S/2}^{S/2}J(x,t)\,dx$, corresponding to the solution in (a). \label{F2a}}
\end{figure}

At the HB point, the newly formed stable time-periodic solution has a finite period, as it can be seen from Fig.\,\ref{F2}. In order to illustrate the dynamics of the depinning of the stationary solution, we set $\alpha=2.08$ and plot in Fig.\,\ref{F2a}(a) snapshots of the time-periodic solution $\rho(x,t)$ at three subsequent times, $t_0$, $t_0+\tau/3$, and $t_0+2\tau /3$, where $t_0$ was chosen as described below and $\tau$ is the temporal period of the solution. Inspection of the density profiles indicates that the depinned solution can be seen as a superposition of two parts: a stationary part with spatial period $L$ and a time-periodic part with spatial period $S=3L$, that slides `on top' of the stationary part. The time-periodic part corresponds to a wave traveling to the left, that is, in the direction of our negative constant drive. Here, at $t=t_0$, the absolute maximum of the density profile is located at the rightmost minimum of the channel potential, $U(x)$. After one third of the temporal period $\tau$, the absolute maximum has moved to the central well of the channel and after two thirds of $\tau$, the maximum has finally reached the leftmost well. After one full period $\tau$, the cycle is repeated. 

The time-periodic solution changes its character along the branch in a continuous manner. With increasing attraction strength the amplitude of the time-periodic part becomes larger as compared to the steady part until finally most of the particles travel. They travel, however, not in the form of a translation of a compact cluster, but rather in the form of a volume transfer of the cluster from one potential well to the next. The temporal period becomes larger with increasing $\alpha$ and the overall flux oscillates between a low absolute value (when the cluster sits in a well) and a large absolute value (when the cluster is transferred to the next well). This is shown in Fig.~\ref{F2a}(b).  With increasing $\alpha$ the dependence of the flux on time becomes increasingly non-harmonic as the cluster spends an increasing fraction of the time period around the three maxima of the channel potential. In the vicinity of the homoclinic bifurcation the density profiles for clusters mainly localised at one of the three maxima closely resemble the corresponding profiles on the three unstable stationary $3L$-periodic solutions.  This also implies that at the homoclinic bifurcation the stable cycle collides with all three unstable equilibria at once.

Summarizing the results displayed in Figs.\,\ref{F2} and \ref{F2a}, we conclude that, for values of $\alpha$ between the points labeled by LP and HB, there exist two stable stationary solutions, with spatial period $L$ and $3L$, respectively. Moreover, between the points HB and hm, a stable time-periodic solution coexists with the stable stationary $3L$-periodic solutions. 
By perturbing the time-periodic density profile with a finite amplitude disturbance, one can induce the transition to the stable stationary $3L$-periodic solution. To do so, one starts a simulation in time with a stable time-periodic solution and adds a finite (mass-conserving) perturbation. If the perturbation is large enough, the solution evolves after a short transient toward the stable 3L periodic stationary solution. Note also that we were not able to find the opposite transition: Perturbing the stable stationary $3L$-periodic solution by shifting it slightly in the direction of the drive will `depin' the cluster only for a short transient. It moves to the left and settles into the next potential well, i.e., it moves to the  stable stationary $3L$-periodic branch that is related by the translation $x\to x-L$.

\subsection{Finite rod length and short range attraction}
\label{finitelength}
\begin{figure*}[t]
\centering
\includegraphics[width=0.9\textwidth]{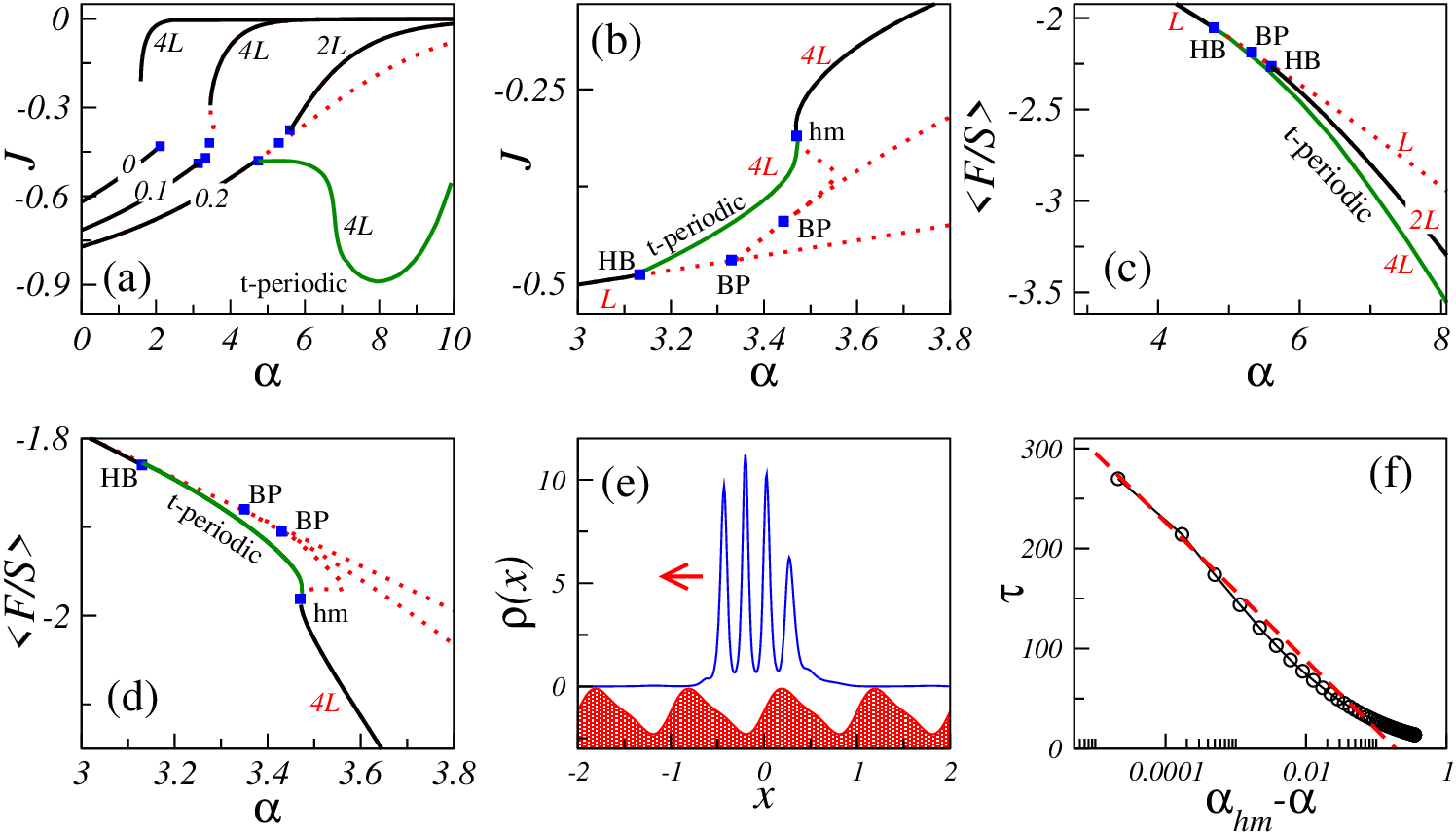}
\caption{(Color online) (a) Current $J$ versus $\alpha$ for $h=0$, 0.1 and 0.2, as indicated on the curves. The other system parameters are $T=1$, $\lambda=5$, $S=4L$, $A=-1$ and $\bar\rho=1$. The symbols denote the Hopf bifurcation (HB) and the branching points (BP). The primary bifurcation is always a HB. The solid and dotted lines for $h=0.1$ and $h=0.2$ represent stable and unstable solution branches, respectively. For $h=0.2$, the branch originating from the HB point is the branch of stable time-periodic solutions. Next to each stable branch is a label indicating the spatial periodicity of the corresponding solutions. (b) displays a magnification of the region in the vicinity of the HB point for $h=0.1$. The branch of stable time-periodic solutions starts at the HB point and terminates at the hm point. Panel (d) shows the average free energy $\langle F/S \rangle$ for $h=0.1$ and the same range of values of $\alpha$ as in (b). Panel (c) shows the free energy of the various solution branches, for $h=0.2$. Panel (e) shows a snapshot of a typical time-periodic solution, obtained for parameters as in (c) and $\alpha=8$. Panel (f) represents the temporal period of the stable time-periodic solutions in (b) as a function of the distance from the hm point, i.e. $(\alpha_\mathrm{hm}-\alpha)$. Dashed line is the decay law $\tau \sim \ln{(\alpha_\mathrm{hm}-\alpha)}$. \label{F4a}}
\end{figure*}

We discuss now the effects of having a finite rod length in addition to the attraction between the particles. In Fig.\,\ref{F4a}(a) we display the bifurcation diagram in terms of the stationary current $J$ as a function of $\alpha$ for rod lengths $h=0, 0.1$, and $0.2$ for a domain length $S=4L$. A relatively small change in the size of the rods is sufficient to cause a significant change in the current. First, one observes that the magnitude of the current at $\alpha=0$ increases with $h$; this also remains true for $\alpha>0$. Second, the critical interaction strength, $\alpha_c$, at which the $1L$-periodic solution looses its stability via a Hopf bifurcation, increases with $h$; i.e.\ as expected a system of finite length rods is more stable than the reference system with $h \to 0$. Third, the region in parameter space in which the stationary $1L$-periodic and the $4L$-periodic solutions coexist, shrinks as $h$ is increased.

This can be explained as follows: For the $4L$-periodic solution to be stable at relatively small values of $\alpha$, one must squeeze all the particles (there are $4$ particles in the system with length $S=4L$ and $\bar\rho=1$) into a small part of the total system, not larger than half a ratchet period, $4h < L/2$. As a consequence, the critical rod length above which the $4L$-periodic and the $1L$-periodic solutions are unlikely to coexist, is approximately $h=0.125$, for $L=1$. A magnification of the bifurcation diagram for $h=0.1$ slightly below this critical value is displayed in Fig.\,\ref{F4a}(b).
%
%
%
There, the stationary $1L$-periodic solutions become unstable at the Hopf bifurcation (HB) and a stable branch of time-periodic density profiles emerges supercritically [heavy green solid line in Fig.\,\ref{F4a}(b)]. Slightly beyond the Hopf bifurcation, the unstable branch of stationary $1L$-periodic solutions undergoes a supercritical period-doubling pitchfork bifurcation (BP) $(\alpha\approx 3.35)$. The emerging branch of stationary $2L$-periodic solutions is unstable w.r.t.\ two modes. It becomes more unstable at a secondary period-doubling pitchfork bifurcation at $(\alpha\approx 3.45)$ (BP). The bifurcating branch consists  of stationary unstable $4L$-periodic solutions with 2 unstable eigenmodes.  One of them is stabilized at a first saddle-node bifurcation at $\alpha\approx 3.55$ where the branch turns back toward smaller $\alpha$. The branch of stationary $4L$-periodic solutions finally becomes stable at another saddle-node bifurcation at $\alpha\approx 3.47$, where it turns again towards larger $\alpha$. The
branch of stable time-periodic solutions terminates as in the case of $h=0$ length rods in a homoclinic bifurcation on the branch of unstable stationary $4L$-periodic solutions. The exact location of the homoclinic bifurcation (labeled ``hm'') is very close to (but numerically clearly distinguished from) the saddle-node bifurcation. The temporal period of the time-periodic solutions diverges logorithmically on approaching the ``hm'' point, as shown in Fig.\,\ref{F4a}(f).

To obtain some indication as to which stable solution might be selected in time evolutions of the DDFT, starting from various initial states, we compute the (time-averaged) Helmholtz free energy per unit length, $\langle F/S \rangle$, for all stable solutions. They are displayed in Fig.\,\ref{F4a}(c) and (d) for $h=0.2$ and $h=0.1$, respectively. In calculating these, we subtract the non-periodic potential energy term, $\int_{-S/2}^{S/2} A \rho(x)\,dx$, associated with the DC drive, from the full expression in Eq.\,(\ref{dft_energy}). This ensures that solutions on branches that are related by the discrete $x\to x+L$ translational symmetry have an identical value for the free energy under periodic BC.


For $h=0.2$, we observe in Fig.\,\ref{F4a}(a) that there exists no branch of stationary $4L$-periodic solutions; instead the stable branch of $2L$-periodic solutions continues toward large $\alpha$. Note that this branch is unstable when it bifurcates from the $1L$ solutions, but becomes stable as a result of a another Hopf bifurcation at $\alpha\approx 5.6$.  The emerging time-periodic branch is unstable and will not be further considered here. The only stable solutions with spatial period equal to the system size, $S=4L$, are the time-periodic ones, which correspond along most of the branch to a single compact cluster of particles traveling in the direction of the drive. Close to the Hopf bifurcation, it resembles a small amplitude wave moving `on top' of the stationary $1L$ state.  Further away from the bifurcation the behaviour resembles the one described above in connection with Fig.~\ref{F2a}: Most of the particles travel in the form of a volume transfer of the cluster from one potential well to the next. Increasing $\alpha$ further, at about $\alpha\approx7$ the flux increases by about 50\% over a very small $\alpha$-range. And the cluster morphology also changes from a compact ``drop-like'' shape to a multi-hump localized structure as depicted in Fig.\,\ref{F4a}(e), with an arrow indicating the direction of motion of the cluster. Each hump corresponds to a single particle. The particles in the cluster are strongly bound together and the distance between the particles remains almost constant as the cluster moves through the system as a single unit. This implies that at $\alpha\approx7$ the transport mode also changes from a volume transfer mode or to a translation mode.

In contrast to the case $h=0.1$, the time-periodic branch continues toward large $\alpha$. In other words, for $h=0.2$ all $4L$-periodic solutions are depinned. In Fig.\,\ref{F4a}(c) we see that time-periodic solutions have on average a lower free energy than the stable stationary $2L$-periodic solutions. However, as the system is permanently out of equilibrium, in general, the solution of lower free energy is not necessarily the one that the system converges to in the long time $t \to \infty$ limit. 
  Thus, for $h=0.2$ the onset of the time-periodic solutions of the DDFT equation is associated with a transition between two major transport modes: (i) At small values of the attraction strength $\alpha$ or, equivalently, for high temperatures, stationary density distributions exist, with the particles uniformly distributed among the wells of the channel potential. Under the action of the stochastic (thermal) noise, the particles jump occasionally either to the right or to the left, but with a higher probability for jumps in the direction of the applied drive. One may call this the ``stationary mode''. (ii) At larger $\alpha$ (or smaller temperature), time-periodic density profiles seem to dominate.  They either correspond to transport from well to well by a volume transfer mode or by a translation mode. The latter
 correspond to depinned compact clusters in which strongly attracting particles travel together. One may call this the ``condensed traveling mode''.  Such a traveling cluster has a characteristic length $\sim hN$ and, in the limit where the attraction $\alpha$ is strong (i.e.\ when $\alpha \gg T$), it moves as a whole in the direction of the drive.

Fig.\,\ref{F4a}(a) shows that the magnitude of the average particle current $J$ is substantially larger (at the same $\alpha$) when transport occurs through the condensed traveling mode, than when in the stationary mode. This can be understood by noticing that for well separated particles, which are effectively not interacting, the average drifting motion of the particles is only resisted by the periodic channel potential. However, when $N$ particles are clustered (bonded) together, then the total pinning force exerted by the channel walls on the cluster is $f=-\sum_{i=1}^N dU(x_i)/d x_i$. As we show in detail below, the value of this net force is very sensitive to the cluster size, and when the length of the cluster $hN$ is an integer multiple of the period of the channel potential $L$, the total pinning force on the cluster vanishes, leading to a maximal drift velocity equal to $A$ \cite{PA10}. 

\section{Low frequency AC drive}
\label{acdrive}
In this section we discuss the behavior of the system when driven by an unbiased AC (square-wave) drive $A(t)=A\,{\rm sgn}\left[ \cos{(\omega t)} \right]$ in the low frequency limit, i.e.\ in the limit $\omega \rightarrow 0$. We focus in particular on the behavior of the average rectification current $\langle J \rangle $. For vanishingly small frequencies, $\langle J \rangle$ is obtained as the arithmetic average of the two unidirectional currents $J^+$ and $J^-$, with $J^{\pm}$ denoting the average currents induced by the DC drives $\pm A$.
\subsection{Maximization of the rectification current}
\label{currmax}
\begin{figure}[t]
\centering
\includegraphics[width=0.48\textwidth]{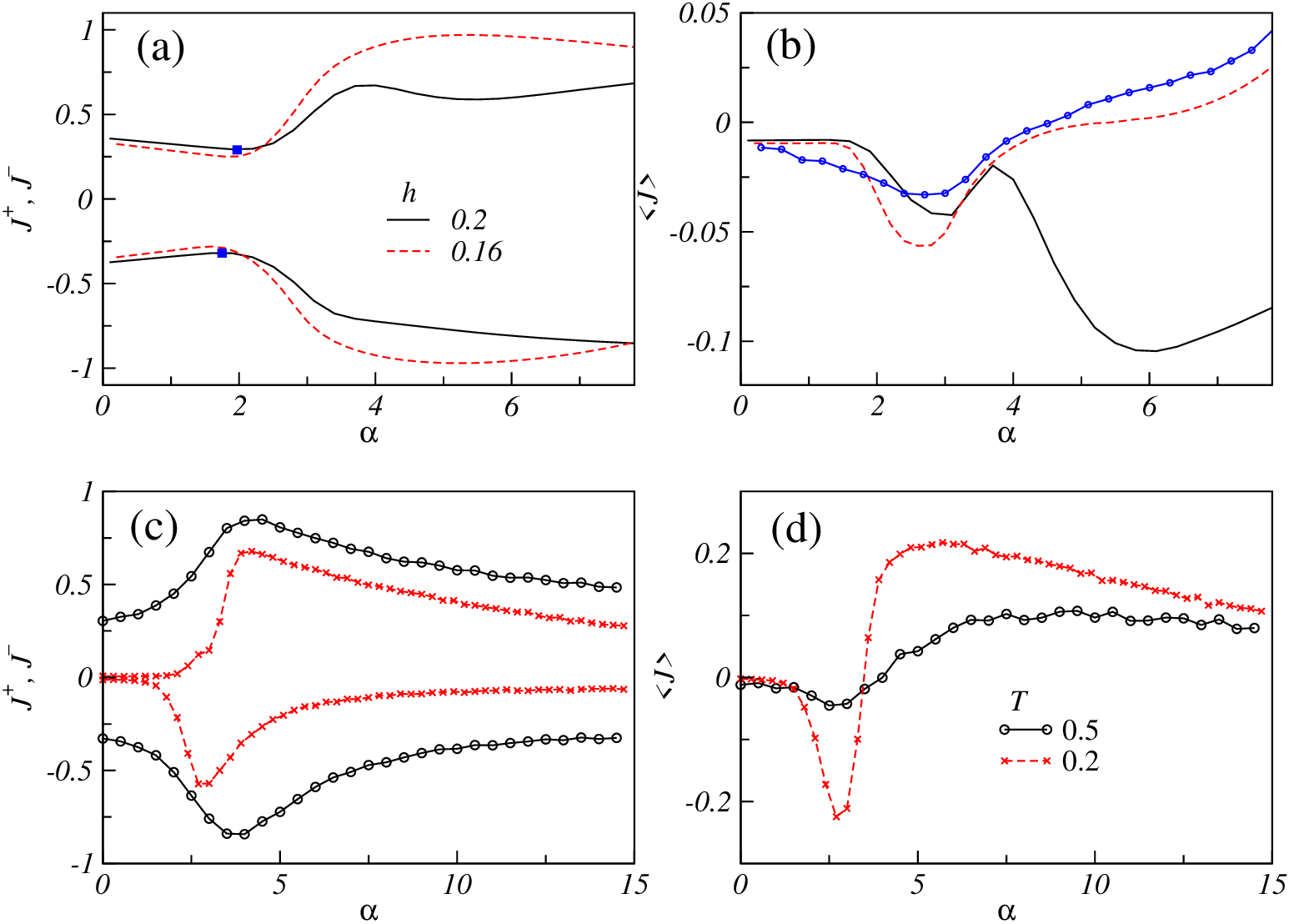}
\caption{(Color online) (a)  The unidirectional currents $J^{\pm}$ as functions of $\alpha$, calculated using the DDFT, for $S=5L$, $N=5$, $A=\pm 1$, $\lambda=3$, $T=0.5$, and for $h=0.2$ (solid line) and $h=0.16$ (dashed line). Symbols ``Solid squares'' mark the corresponding Hopf bifurcation of the stationary density distribution for $h=0.2$. Note that to left of the Hopf bifurcation we show the current for the stationary solutions and to the right for the time-periodic solution.
(b) The rectification current $\langle J \rangle$ versus $\alpha$, computed as the arithmetic mean of $J^+$  and $J^-$ in (a) (curves) and from direct Brownian dynamics simulations with $h_{\rm eff}=0.16$ (symbols).  All other system parameters are as in (a). Panel (c) gives the unidirectional currents $J^{\pm}$ obtained from Brownian dynamics simulations as functions of $\alpha$ for $T=0.5$ (solid line) and $T=0.2$ (dashed line). The remaining parameters are $S=10$, $N=10$, $A=\pm 1$, $\lambda=3$, and $h_{\rm eff}=0.16$. (d) The rectification currents $\langle J \rangle$, computed as the arithmetic mean of the currents $J^+$  and $J^-$ in (c). \label{F5}}
\end{figure}
As shown in the previous section, for constant drive $A$, increasing the pair attraction strength $\alpha$, leads to the onset of a condensed traveling transport mode associated with the clustering of the particles traveling in the direction of the drive. As the condensation sets in, the opposite unidirectional currents $J^{\pm}$ increase in magnitude. However, due to the asymmetry of the channel potential, the condensation sets in at different values of $\alpha$, depending on the orientation of the drive. This phenomenon is illustrated in Fig.\,\ref{F5}(a), where the two relevant HB points are marked, for the case when $h=0.2$. Owing to the spatial asymmetry of $U(x)$, the depinning of the stationary density profile when the drive is $-A$, with current $J^{-}$, occurs at a lower value of $\alpha$ than when the drive is $+A$, with current $J^{+}$. Therefore, when $\alpha$ is gradually increased beyond the value at the HB point for negative drive $-A$, the cycle averaged rectification current $\langle J \rangle = (1/2)(J^{+}+J^{-})$, is negative and increases in absolute value, as shown in Fig.\,\ref{F5}(b). As $\alpha$ is further increased to the value at the HB point for positive drive $+A$, the magnitude of $\langle J \rangle$ reaches a local maximum as a function of $\alpha$. Increasing $\alpha$ even further results in a decrease in the magnitude of $\langle J \rangle$. This occurs because the particles are now transported as a condensed traveling mode in both directions.

A qualitatively similar behavior of $\langle J \rangle$ is found for a range of different values of the particle size $h$. However, on increasing $\alpha$ even further, so that it is well above the value at the HB points, the dependence of  $\langle J \rangle$ on $\alpha$ becomes very sensitive to the value of $h$. For instance, in Fig.\,\ref{F5}(b) the rectification current attains a second minimum at around $\alpha=6$ for $h=0.2$, whereas for $h=0.16$ the second minimum disappears and $\langle J \rangle$ increases monotonically as a function of $\alpha$.

To confirm the validity of the (mean field) DDFT results, we performed Brownian dynamics computer simulations -- i.e.\ we numerically integrated the Langevin equations of motion (\ref{langevin}), in order to compare with our DDFT results. In order to make the simulations more convenient to implement, we replace the hard core potential, $w_{\rm hr}$, by an equivalent, more tractable soft core potential, $w_{\rm s}(x_{ij}) = \epsilon (h_*/x_{ij})^{19}$, where the constants $\epsilon$ and $h_*$ can be tuned to reproduce the desired effective hard-core length of the potential. For fixed $\epsilon$ and $h_*$ the effective hard core length $h_{\rm eff}$ of the particles becomes a function of $\alpha$, $T$ and, in general, also of the number of particles $N$ \cite{barker76}. In our simulations we set $\epsilon=0.01$ and $h_*=0.2$ which corresponds to an effective hard-core $h_{\rm eff}\approx 0.16$, for $\alpha=10$. Our numerical data suggests that the dependence of $h_{\rm eff}$ on $T$ and  $N$ is rather weak, and can therefore be neglected.

In Fig.\,\ref{F5}(b) we compare the DDFT predictions for $h=0.16$ with the corresponding simulation results for $h_{\rm eff}\approx 0.16$. The first minimum in the current $\langle J \rangle$ as a function of $\alpha$ is clearly confirmed by the Brownian dynamics simulation results for $N=5$ particles. The simulation results displayed in Figs.\,\ref{F5}(c) and (d) also show that the overall structure of $\langle J \rangle$ as a function of $\alpha$ does not change much as $N$ is increased up to $10$. Moreover, as the temperature is decreased from $T=0.5$ down to $T=0.2$, the maximum in the magnitude of the current curve, $|\langle J \rangle|$, becomes even more pronounced, with the magnitude of the peak rectification current increasing by one order of magnitude. This effect, which is well established in the ratchet literature \cite{RMP09}, underlines the key role of noise in activating transport (in either direction) when the amplitude of the drive is smaller than both the depinning thresholds, $A_L$ and $A_R$, of the ratchet potential, $U(x)$.
\subsection{Strong attraction limit}
\label{strongatt}
\begin{figure}[t]
\centering
\includegraphics[width=0.48\textwidth]{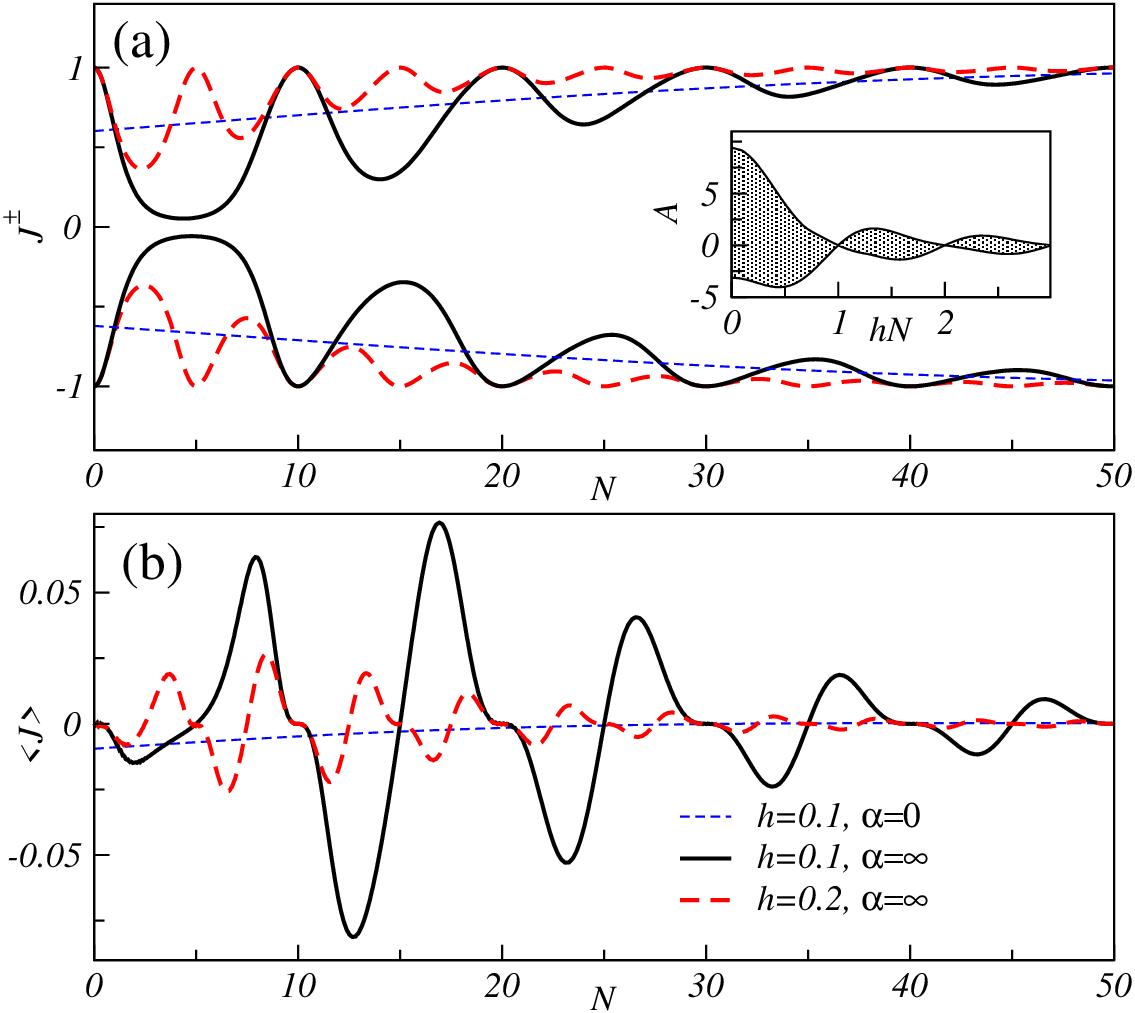}
\caption{(Color online)  (a) The unidirectional currents $J^{\pm}$ as functions of the number of particles $N$, in a system with total length $S=10L$. The heavy solid and dashed curves correspond to the currents in the limit of very strong attraction ($\alpha\to\infty$) for $h=0.1$ and $h=0.2$, respectively. The thin dashed curves represent the currents for non-attracting particles ($\alpha=0$) with $h=0.1$. In the inset we display the $T=0$ limiting values for the critical amplitude $A$ required for a current to flow, as a function of $hN$. In (b) we display the rectification currents $\langle J\rangle$ obtained from the currents displayed in (a).  \label{F6}}
\end{figure}
In order to study the properties of the system when the attraction between the particles dominates over the thermal motion of the particles and the pinning by the external potential, we consider the limit $\alpha \to \infty$. This allows us to reduce the system of equations (\ref{langevin}) to a single equation of motion for the center of mass of the particle condensate, $y=(1/N)\sum_{i=1}^{N} x_i$. As noted above in Sec.\ \ref{finitelength}, the total force exerted by the channel potential on the condensate is $f=-\sum_{i=1}^{N}dU(x_i)/dx_i$. If we assume that the pair attraction is so strong that the rods are closely packed together in a single condensate with their ends touching, the total force $f$ can be rewritten as $f=-\sum_{i=1}^{N}dU(x+(i-1)h)/dx$, where $x$ denotes the coordinate of the center of the first particle in the file. Now, if we assume that $h$ is small compared to the period $L$ of the channel potential, then the sum can be replaced by an integral:
\begin{eqnarray}
f \approx -\frac{1}{hN}\int_{x}^{x+hN}\frac{dU(y)}{dy}dy \notag \\
= -\frac{U(x+hN)-U(x)}{hN},
\end{eqnarray}
leading to the following effective equation of motion for  the center of mass:
\begin{eqnarray}
\label{langevin2}
\frac{d x}{d t} = -\frac{U(x+hN)-U(x)}{hN} +A(t)+ \sqrt{\frac{2T}{N}}\xi(t).
\end{eqnarray}
Here, $\xi(t)$ has the same statistics as $\xi_i(t)$ in Eq.\,(\ref{langevin}). To derive Eq.\,(\ref{langevin2}), we use the fact that the sum of $N$ independent sources of Gaussian white noise with variance $1$ is also a Gaussian noise, but with variance $1/N$.

Equation (\ref{langevin2}) corresponds to the equation of motion for a single Brownian particle diffusing in the effective external potential $V_{\rm eff}(x) = \frac{1}{hN} \int [U(x+hN)-U(x)]\,dx$, in contact with a thermal bath with temperature $T/N$. The first observation from Eq.\ \eqref{langevin2} is that for large condensates, diffusion becomes negligible, so that $J^{\pm}$ become sizable only if the drive amplitude, $A$, overcomes the pinning force induced by the effective potential $V_{\rm eff}(x)$. For $T\equiv 0$, this critical amplitude $A$ is plotted in the inset of Fig.\,\ref{F6}(a), as a function of the size of the condensate $hN$. Within the shaded area, the condensate is pinned by the effective external potential $V_{\rm eff}$; depinning occurs either to the left or to the right, depending on the drive orientation. Note that for $hN \to 0$, the right and left critical amplitudes coincide with the single particle depinning thresholds, $A_{L}$ and $A_R$, introduced in Sec. \ref{DC}. Similarly to the case for pointlike particles \cite{sav04}, selecting an appropriate combinations of $h$ and $N$, one can achieve the complete locking of the condensed mode in one direction, but not in the other \cite{PA10}, which yields the upper bound $|A|/2$ for the modulus of $\langle J \rangle$.

Finally, using Eq.\,(\ref{langevin2}), we compare in Fig.\ref{F6} the efficiency of the low frequency transport of strongly attracting ($\alpha \to \infty$) and non-interacting ($\alpha =0$) particles. We fix the size of the particles $h$ and change the average density $\bar \rho$ by changing the number of particles $N$ in the system. The unidirectional currents for the condensate oscillate with $N$ and hit the respective upper (lower) bound, $J^{\pm}=\pm |A|$, for $hN$ equal to a multiple of $L$. In the absence of particle attraction, $|J^{\pm}|$ increases monotonically with $N$ and attain the same upper bound only for $hN=S$. The corresponding rectification currents are shown in Fig.\ref{F6}(b). For certain combinations of $h$ and $N$, the magnitude of the current of the condensate is several orders of magnitude larger than for non-attracting particles. 

\section{Concluding remarks}
\label{conc}
In this paper we have developed a DDFT for studying the dynamics of a file of attracting colloidal particles confined within a channel that exerts a periodic ratchet potential on the colloids. We find that the attraction between the colloids leads to rather rich behavior in the DDFT model when the particles are driven, including transitions from stationary to time-periodic density profiles as the strength of the attraction between the particles is increased. We also find that for strong enough attraction, there can be coexistence of stable stationary density profiles with different spatial periods and time-periodic density profiles, each with different values for the particle current $J$.

These dynamical transitions in our model stem from the fact that the approximate free energy functional \eqref{dft_energy} on which our DDFT is based, predicts that the system exhibits gas-liquid phase separation for sufficiently large values of the ratio $\alpha/Th\lambda$. This prediction comes as a consequence of the mean-field approximation made in constructing the free energy. In reality, for a system containing a finite number of particles, there is no true phase transition. Furthermore, since the system is one-dimensional, there is no phase transition even in the infinite sized system (i.e\ in the thermodynamic limit when $N,S\to\infty$, with average density $\bar{\rho}=N/S$ remaining constant). In 1D systems such as that studied here, as the attraction strength $\alpha$ is increased, the particles increasingly tend to gather together, but no true phase transition can be defined. Thus, in reality, as can be inferred from our Brownian dynamics simulation results, there are no `sharp' transitions from the pinned to the depinned (time-periodic) state, as $\alpha$ is increased. Thus, we expect that fluctuations will round the predicted transitions. Nonetheless, as the comparison with the Brownian dynamics simulations show, the results from our DDFT do capture the main features of the system - i.e.\ that for lower values of the attraction strength $\alpha$, the particles are uniformly distributed and that at higher values of $\alpha$ the particles gather to form a cluster, and that if the system length $S$ is sufficiently long, this clustering leads to time-periodic currents $J$ when the system is driven.

In our discussions above we have pointed out that similarities exist between the DDFT equation \eqref{eq3} for the particle density employed here and thin film equations that are used to model the dynamics of films and drops of partially wetting liquids on heterogenous solid substrates with and without additional driving forces \cite{KaTh07}. 
The similarities result from the fact that in both cases kinetic equations for conserved fields are used, and that the respective free energy functionals contain terms which result in similar physical effects. For instance, the role of the particle-particle interactions in the present work is taken by wettability effects in the context of droplet dynamics. The parallels between the two systems have allowed us to use the knowledge gained from studying one system to understand aspects of the other. In particular, in the present work we have drawn on the understanding of depinning mechanisms developed for thin films in Refs.~\cite{dep2,BKHT11}. Furthermore, Ref.~\cite{BKHT11} indicates that one might encounter rich nonlinear behaviour when considering the behaviour of attractive hard-core particles in wider corrugated channels, i.e., without the `restriction' of single file motion. Note, however, that there are clear limits to the similarities: The thin film models referred to above do not account for any effect that is equivalent to the freezing 
instability discussed above. We believe that studying in detail the similarities and differences between DDFT and thin film models is worthwhile, as it will allow for much cross fertilisation of ideas and techniques between the two fields.

One of the most striking features of our system is that the current $J$ depends very sensitively on the size of the particles $h$ and on the total number of particles in the system $N$, particularly when the particles are strongly attracted to one another so that they are bound together to form a cluster that moves as a unit through the system, when there is an external drive on the system. In fact, the direction of travel can be completely reversed when the system is driven by an AC potential, simply by changing the number of particles in the file by one -- i.e.\ adding an extra particle to a file can cause it to reverse its direction of motion without changing the external drive. This means that one can use the present system to form a molecular shuttle that moves back and forth between two docking stations, loading and unloading single particles from a source to a sink docking station \cite{PA10}. As the process repeats, a steady flux is established along the channel. This mechanism can be highly efficient if the system parameters are carefully tuned.

The present model thus provides a useful system for developing a deep understanding of the behavior of driven macromolecular and colloidal systems occurring in nanoscience and biology. In particular, by using DDFT, which is based on a fully microscopic expression for the Helmholtz free energy functional \eqref{dft_energy}, we are able to build into our theory a reliable description of the correlations between the particles, and their influence on the dynamics of the system as a whole.

\section*{Acknowledgements}

This work was partly supported by the HPC-Europa2 Transnational Access Programme, proposal No. 278. AJA gratefully acknowledges support from RCUK. FM acknowledges partial support from the Seventh Framework Programme under grant agreement No. 256959, project NANOPOWER.



\begin{thebibliography}{48}
\expandafter\ifx\csname natexlab\endcsname\relax\def\natexlab#1{#1}\fi
\expandafter\ifx\csname bibnamefont\endcsname\relax
  \def\bibnamefont#1{#1}\fi
\expandafter\ifx\csname bibfnamefont\endcsname\relax
  \def\bibfnamefont#1{#1}\fi
\expandafter\ifx\csname citenamefont\endcsname\relax
  \def\citenamefont#1{#1}\fi
\expandafter\ifx\csname url\endcsname\relax
  \def\url#1{\texttt{#1}}\fi
\expandafter\ifx\csname urlprefix\endcsname\relax\def\urlprefix{URL }\fi
\providecommand{\bibinfo}[2]{#2}
\providecommand{\eprint}[2][]{\url{#2}}

\bibitem[{\citenamefont{H{\"a}nggi and Marchesoni}(2009)}]{RMP09}
\bibinfo{author}{\bibfnamefont{P.}~\bibnamefont{H{\"a}nggi}} \bibnamefont{and}
  \bibinfo{author}{\bibfnamefont{F.}~\bibnamefont{Marchesoni}},
  \bibinfo{journal}{Rev. Mod. Phys.} \textbf{\bibinfo{volume}{81}},
  \bibinfo{pages}{387} (\bibinfo{year}{2009}).

\bibitem[{\citenamefont{Hille}(2001)}]{hille01}
\bibinfo{author}{\bibfnamefont{B.}~\bibnamefont{Hille}},
  \emph{\bibinfo{title}{Channels of Excitable Membranes}}
  (\bibinfo{publisher}{Sinauer Asc., Sunderland}, \bibinfo{year}{2001}).

\bibitem[{\citenamefont{K{\"a}rger and Ruthven}(1992)}]{kaerger92}
\bibinfo{author}{\bibfnamefont{J.}~\bibnamefont{K{\"a}rger}} \bibnamefont{and}
  \bibinfo{author}{\bibfnamefont{D.~M.} \bibnamefont{Ruthven}},
  \emph{\bibinfo{title}{Diffusion in Zeolites and Other Microporous Solids}}
  (\bibinfo{publisher}{Wiley, New York}, \bibinfo{year}{1992}).

\bibitem[{\citenamefont{Squires and Quake}(2005)}]{squires05}
\bibinfo{author}{\bibfnamefont{T.~M.} \bibnamefont{Squires}} \bibnamefont{and}
  \bibinfo{author}{\bibfnamefont{S.~R.} \bibnamefont{Quake}},
  \bibinfo{journal}{Rev. Mod. Phys.} \textbf{\bibinfo{volume}{77}},
  \bibinfo{pages}{977} (\bibinfo{year}{2005}).

\bibitem[{\citenamefont{Wei et~al.}(2000)\citenamefont{Wei, Bechinger, and
  Leiderer}}]{wei00}
\bibinfo{author}{\bibfnamefont{Q.~H.} \bibnamefont{Wei}},
  \bibinfo{author}{\bibfnamefont{C.}~\bibnamefont{Bechinger}},
  \bibnamefont{and} \bibinfo{author}{\bibfnamefont{P.}~\bibnamefont{Leiderer}},
  \bibinfo{journal}{Science} \textbf{\bibinfo{volume}{287}},
  \bibinfo{pages}{625} (\bibinfo{year}{2000}).

\bibitem[{\citenamefont{Lutz et~al.}(2004)\citenamefont{Lutz, Kollmann, and
  Bechinger}}]{lutz04}
\bibinfo{author}{\bibfnamefont{C.}~\bibnamefont{Lutz}},
  \bibinfo{author}{\bibfnamefont{M.}~\bibnamefont{Kollmann}}, \bibnamefont{and}
  \bibinfo{author}{\bibfnamefont{C.}~\bibnamefont{Bechinger}},
  \bibinfo{journal}{Phys. Rev. Lett.} \textbf{\bibinfo{volume}{93}},
  \bibinfo{pages}{026001} (\bibinfo{year}{2004}).

\bibitem[{\citenamefont{Besseling et~al.}(1998)\citenamefont{Besseling,
  Niggebrugge, and Kes}}]{besseling98}
\bibinfo{author}{\bibfnamefont{R.}~\bibnamefont{Besseling}},
  \bibinfo{author}{\bibfnamefont{R.}~\bibnamefont{Niggebrugge}},
  \bibnamefont{and} \bibinfo{author}{\bibfnamefont{P.~H.} \bibnamefont{Kes}},
  \bibinfo{journal}{Phys. Rev. Lett.} \textbf{\bibinfo{volume}{82}},
  \bibinfo{pages}{3144} (\bibinfo{year}{1998}).

\bibitem[{\citenamefont{Coupier et~al.}(2007)\citenamefont{Coupier, Sain~Jean,
  and Guthmann}}]{coupier07}
\bibinfo{author}{\bibfnamefont{G.}~\bibnamefont{Coupier}},
  \bibinfo{author}{\bibfnamefont{M.}~\bibnamefont{Sain~Jean}},
  \bibnamefont{and} \bibinfo{author}{\bibfnamefont{C.}~\bibnamefont{Guthmann}},
  \bibinfo{journal}{Europhysics Lett.} \textbf{\bibinfo{volume}{77}},
  \bibinfo{pages}{60001} (\bibinfo{year}{2007}).

\bibitem[{\citenamefont{Ashkin et~al.}(1990)\citenamefont{Ashkin, Sch{\"u}tze,
  Dziedzie, Euteneuer, and Schliwa}}]{ashkin90}
\bibinfo{author}{\bibfnamefont{A.}~\bibnamefont{Ashkin}},
  \bibinfo{author}{\bibfnamefont{K.}~\bibnamefont{Sch{\"u}tze}},
  \bibinfo{author}{\bibfnamefont{j.~M.} \bibnamefont{Dziedzie}},
  \bibinfo{author}{\bibfnamefont{U.}~\bibnamefont{Euteneuer}},
  \bibnamefont{and} \bibinfo{author}{\bibfnamefont{M.}~\bibnamefont{Schliwa}},
  \bibinfo{journal}{Nature (London)} \textbf{\bibinfo{volume}{348}},
  \bibinfo{pages}{346} (\bibinfo{year}{1990}).

\bibitem[{\citenamefont{Wambaugh et~al.}(1999)\citenamefont{Wambaugh,
  Reichhardt, Olson, Marchesoni, and Nori}}]{wambaugh}
\bibinfo{author}{\bibfnamefont{J.~F.} \bibnamefont{Wambaugh}},
  \bibinfo{author}{\bibfnamefont{C.}~\bibnamefont{Reichhardt}},
  \bibinfo{author}{\bibfnamefont{C.~J.} \bibnamefont{Olson}},
  \bibinfo{author}{\bibfnamefont{F.}~\bibnamefont{Marchesoni}},
  \bibnamefont{and} \bibinfo{author}{\bibfnamefont{F.}~\bibnamefont{Nori}},
  \bibinfo{journal}{Phys. Rev. Lett.} \textbf{\bibinfo{volume}{83}},
  \bibinfo{pages}{5106–5109} (\bibinfo{year}{1999}).

\bibitem[{\citenamefont{Taloni and Marchesoni}(2006)}]{taloni06}
\bibinfo{author}{\bibfnamefont{A.}~\bibnamefont{Taloni}} \bibnamefont{and}
  \bibinfo{author}{\bibfnamefont{F.}~\bibnamefont{Marchesoni}},
  \bibinfo{journal}{Phys. Rev. Lett.} \textbf{\bibinfo{volume}{96}},
  \bibinfo{pages}{020601} (\bibinfo{year}{2006}).

\bibitem[{\citenamefont{Derenyi and Vicsek}(1995)}]{vicsek95}
\bibinfo{author}{\bibfnamefont{I.}~\bibnamefont{Derenyi}} \bibnamefont{and}
  \bibinfo{author}{\bibfnamefont{T.}~\bibnamefont{Vicsek}},
  \bibinfo{journal}{Phys. Rev. Lett.} \textbf{\bibinfo{volume}{75}},
  \bibinfo{pages}{374} (\bibinfo{year}{1995}).

\bibitem[{\citenamefont{Barrat and Hansen}(2003)}]{BarratHansen2003}
\bibinfo{author}{\bibfnamefont{J.-L.} \bibnamefont{Barrat}} \bibnamefont{and}
  \bibinfo{author}{\bibfnamefont{J.-P.} \bibnamefont{Hansen}},
  \emph{\bibinfo{title}{Basic Concepts for Simple and Complex Liquid}}
  (\bibinfo{publisher}{Cambridge University Press},
  \bibinfo{address}{Cambridge}, \bibinfo{year}{2003}).

\bibitem[{\citenamefont{Hansen and McDonald}(2006)}]{hansen2006tsl}
\bibinfo{author}{\bibfnamefont{J.-P.} \bibnamefont{Hansen}} \bibnamefont{and}
  \bibinfo{author}{\bibfnamefont{I.~R.} \bibnamefont{McDonald}},
  \emph{\bibinfo{title}{Theory of Simple Liquids}}
  (\bibinfo{publisher}{Academic, London}, \bibinfo{year}{2006}).

\bibitem[{\citenamefont{Poon}(2002)}]{poon02}
\bibinfo{author}{\bibfnamefont{W.~C.~K.} \bibnamefont{Poon}},
  \bibinfo{journal}{J. Phys. Condens. Matter} \textbf{\bibinfo{volume}{24}},
  \bibinfo{pages}{R859} (\bibinfo{year}{2002}).

\bibitem[{\citenamefont{Stradner et~al.}(2004)\citenamefont{Stradner, Sedgwick,
  Cardinaux, C.~K.~Poon, Egelhaaf, and Schurtenberger}}]{stradner04}
\bibinfo{author}{\bibfnamefont{A.}~\bibnamefont{Stradner}},
  \bibinfo{author}{\bibfnamefont{H.}~\bibnamefont{Sedgwick}},
  \bibinfo{author}{\bibfnamefont{F.}~\bibnamefont{Cardinaux}},
  \bibinfo{author}{\bibfnamefont{W.}~\bibnamefont{C.~K.~Poon}},
  \bibinfo{author}{\bibfnamefont{S.~U.} \bibnamefont{Egelhaaf}},
  \bibnamefont{and}
  \bibinfo{author}{\bibfnamefont{P.}~\bibnamefont{Schurtenberger}},
  \bibinfo{journal}{Nature} \textbf{\bibinfo{volume}{432}},
  \bibinfo{pages}{492} (\bibinfo{year}{2004}).

\bibitem[{\citenamefont{Savel'ev
  et~al.}(2004{\natexlab{a}})\citenamefont{Savel'ev, Marchesoni, and
  Nori}}]{savsav}
\bibinfo{author}{\bibfnamefont{S.}~\bibnamefont{Savel'ev}},
  \bibinfo{author}{\bibfnamefont{F.}~\bibnamefont{Marchesoni}},
  \bibnamefont{and} \bibinfo{author}{\bibfnamefont{F.}~\bibnamefont{Nori}},
  \bibinfo{journal}{Phys. Rev. Lett.} \textbf{\bibinfo{volume}{92}},
  \bibinfo{pages}{160602} (\bibinfo{year}{2004}{\natexlab{a}}).

\bibitem[{\citenamefont{Sholl and Fichthorn}(1997)}]{sholl97}
\bibinfo{author}{\bibfnamefont{D.~S.} \bibnamefont{Sholl}} \bibnamefont{and}
  \bibinfo{author}{\bibfnamefont{K.~A.} \bibnamefont{Fichthorn}},
  \bibinfo{journal}{Phys. Rev. Lett.} \textbf{\bibinfo{volume}{79}},
  \bibinfo{pages}{3569} (\bibinfo{year}{1997}).

\bibitem[{\citenamefont{Dubbeldam et~al.}(2003)\citenamefont{Dubbeldam, Calero,
  Maesen, and Smit}}]{dubbeldam03}
\bibinfo{author}{\bibfnamefont{D.}~\bibnamefont{Dubbeldam}},
  \bibinfo{author}{\bibfnamefont{S.}~\bibnamefont{Calero}},
  \bibinfo{author}{\bibfnamefont{T.~L.~M.} \bibnamefont{Maesen}},
  \bibnamefont{and} \bibinfo{author}{\bibfnamefont{B.}~\bibnamefont{Smit}},
  \bibinfo{journal}{Phys. Rev. Lett.} \textbf{\bibinfo{volume}{90}},
  \bibinfo{pages}{245901} (\bibinfo{year}{2003}).

\bibitem[{\citenamefont{Smit and Maesen}(2008)}]{smit08}
\bibinfo{author}{\bibfnamefont{B.}~\bibnamefont{Smit}} \bibnamefont{and}
  \bibinfo{author}{\bibfnamefont{T.~L.~M.} \bibnamefont{Maesen}},
  \bibinfo{journal}{Nature} \textbf{\bibinfo{volume}{451}},
  \bibinfo{pages}{06552} (\bibinfo{year}{2008}).

\bibitem[{\citenamefont{Pototsky et~al.}(2010)\citenamefont{Pototsky, Archer,
  Bestehorn, Merkt, Savel'ev, and Marchesoni}}]{PA10}
\bibinfo{author}{\bibfnamefont{A.}~\bibnamefont{Pototsky}},
  \bibinfo{author}{\bibfnamefont{A.~J.} \bibnamefont{Archer}},
  \bibinfo{author}{\bibfnamefont{M.}~\bibnamefont{Bestehorn}},
  \bibinfo{author}{\bibfnamefont{D.}~\bibnamefont{Merkt}},
  \bibinfo{author}{\bibfnamefont{S.}~\bibnamefont{Savel'ev}}, \bibnamefont{and}
  \bibinfo{author}{\bibfnamefont{F.}~\bibnamefont{Marchesoni}},
  \bibinfo{journal}{Phys. Rev. E} \textbf{\bibinfo{volume}{82}},
  \bibinfo{pages}{030401(R)} (\bibinfo{year}{2010}).

\bibitem[{\citenamefont{Marconi and Tarazona}(1999)}]{marini99}
\bibinfo{author}{\bibfnamefont{U.~M.~B.} \bibnamefont{Marconi}}
  \bibnamefont{and} \bibinfo{author}{\bibfnamefont{P.}~\bibnamefont{Tarazona}},
  \bibinfo{journal}{J. Chem Phys.} \textbf{\bibinfo{volume}{110}},
  \bibinfo{pages}{8032} (\bibinfo{year}{1999}).

\bibitem[{\citenamefont{Marconi and Tarazona}(2000)}]{marini00}
\bibinfo{author}{\bibfnamefont{U.~M.~B.} \bibnamefont{Marconi}}
  \bibnamefont{and} \bibinfo{author}{\bibfnamefont{P.}~\bibnamefont{Tarazona}},
  \bibinfo{journal}{J. Phys.: Condens Matter} \textbf{\bibinfo{volume}{12}},
  \bibinfo{pages}{A413} (\bibinfo{year}{2000}).

\bibitem[{\citenamefont{Archer and Evans}(2004)}]{archer04}
\bibinfo{author}{\bibfnamefont{A.~J.} \bibnamefont{Archer}} \bibnamefont{and}
  \bibinfo{author}{\bibfnamefont{R.}~\bibnamefont{Evans}}, \bibinfo{journal}{J.
  Chem. Phys.} \textbf{\bibinfo{volume}{121}}, \bibinfo{pages}{4246}
  (\bibinfo{year}{2004}).

\bibitem[{\citenamefont{Archer and Rauscher}(2000)}]{archer04b}
\bibinfo{author}{\bibfnamefont{A.~J.} \bibnamefont{Archer}} \bibnamefont{and}
  \bibinfo{author}{\bibfnamefont{M.}~\bibnamefont{Rauscher}},
  \bibinfo{journal}{J.Phys. A: Math. Gen.} \textbf{\bibinfo{volume}{37}},
  \bibinfo{pages}{9325} (\bibinfo{year}{2000}).

\bibitem[{\citenamefont{Evans}(1992)}]{evans1992fif}
\bibinfo{author}{\bibfnamefont{R.}~\bibnamefont{Evans}},
  \emph{\bibinfo{title}{Fundamentals of Inhomogeneous Fluids}}
  (\bibinfo{publisher}{Dekker, New York}, \bibinfo{year}{1992}).

\bibitem[{\citenamefont{Evans}(1979)}]{evans}
\bibinfo{author}{\bibfnamefont{R.}~\bibnamefont{Evans}}, \bibinfo{journal}{Adv.
  Phys.} \textbf{\bibinfo{volume}{28}}, \bibinfo{pages}{143}
  (\bibinfo{year}{1979}).

\bibitem[{\citenamefont{Percus}(1978)}]{percus76}
\bibinfo{author}{\bibfnamefont{J.~K.} \bibnamefont{Percus}},
  \bibinfo{journal}{J. Stat. Phys.} \textbf{\bibinfo{volume}{15}},
  \bibinfo{pages}{505} (\bibinfo{year}{1978}).

\bibitem[{\citenamefont{Penna and Tarazona}(2003)}]{penna03}
\bibinfo{author}{\bibfnamefont{F.}~\bibnamefont{Penna}} \bibnamefont{and}
  \bibinfo{author}{\bibfnamefont{P.}~\bibnamefont{Tarazona}},
  \bibinfo{journal}{J. Chem. Phys.} \textbf{\bibinfo{volume}{119}},
  \bibinfo{pages}{1766} (\bibinfo{year}{2003}).

\bibitem[{\citenamefont{Kalliadasis and Thiele}(2007)}]{KaTh07}
\bibinfo{editor}{\bibfnamefont{S.}~\bibnamefont{Kalliadasis}} \bibnamefont{and}
  \bibinfo{editor}{\bibfnamefont{U.}~\bibnamefont{Thiele}}, eds.,
  \emph{\bibinfo{title}{Thin Films of Soft Matter}}
  (\bibinfo{publisher}{Springer}, \bibinfo{address}{Wien / New York},
  \bibinfo{year}{2007}), ISBN \bibinfo{isbn}{978-3-211-69807-5}.

\bibitem[{\citenamefont{Thiele and John}(2010)}]{ThJo10}
\bibinfo{author}{\bibfnamefont{U.}~\bibnamefont{Thiele}} \bibnamefont{and}
  \bibinfo{author}{\bibfnamefont{K.}~\bibnamefont{John}},
  \bibinfo{journal}{Chem. Phys.} \textbf{\bibinfo{volume}{375}},
  \bibinfo{pages}{578} (\bibinfo{year}{2010}).

\bibitem[{\citenamefont{Thiele et~al.}(2003)\citenamefont{Thiele, Brusch,
  Bestehorn, and B{\"a}r}}]{TBBB03}
\bibinfo{author}{\bibfnamefont{U.}~\bibnamefont{Thiele}},
  \bibinfo{author}{\bibfnamefont{L.}~\bibnamefont{Brusch}},
  \bibinfo{author}{\bibfnamefont{M.}~\bibnamefont{Bestehorn}},
  \bibnamefont{and} \bibinfo{author}{\bibfnamefont{M.}~\bibnamefont{B{\"a}r}},
  \bibinfo{journal}{Eur. Phys. J. E} \textbf{\bibinfo{volume}{11}},
  \bibinfo{pages}{255} (\bibinfo{year}{2003}).

\bibitem[{\citenamefont{Thiele and Knobloch}(2006{\natexlab{a}})}]{dep2}
\bibinfo{author}{\bibfnamefont{U.}~\bibnamefont{Thiele}} \bibnamefont{and}
  \bibinfo{author}{\bibfnamefont{E.}~\bibnamefont{Knobloch}},
  \bibinfo{journal}{New J. Phys.} \textbf{\bibinfo{volume}{313}},
  \bibinfo{pages}{1} (\bibinfo{year}{2006}{\natexlab{a}}).

\bibitem[{\citenamefont{de~Gennes}(1985)}]{deGe85}
\bibinfo{author}{\bibfnamefont{P.-G.} \bibnamefont{de~Gennes}},
  \bibinfo{journal}{Rev. Mod. Phys.} \textbf{\bibinfo{volume}{57}},
  \bibinfo{pages}{827} (\bibinfo{year}{1985}).

\bibitem[{\citenamefont{John and Thiele}(2010)}]{JoTh10}
\bibinfo{author}{\bibfnamefont{K.}~\bibnamefont{John}} \bibnamefont{and}
  \bibinfo{author}{\bibfnamefont{U.}~\bibnamefont{Thiele}},
  \bibinfo{journal}{Phys. Rev. Lett.} \textbf{\bibinfo{volume}{104}},
  \bibinfo{pages}{107801} (\bibinfo{year}{2010}).

\bibitem[{\citenamefont{John and Thiele}(2007)}]{JoTh07}
\bibinfo{author}{\bibfnamefont{K.}~\bibnamefont{John}} \bibnamefont{and}
  \bibinfo{author}{\bibfnamefont{U.}~\bibnamefont{Thiele}},
  \bibinfo{journal}{Appl. Phys. Lett.} \textbf{\bibinfo{volume}{90}},
  \bibinfo{pages}{264102} (\bibinfo{year}{2007}).

\bibitem[{\citenamefont{Cross and Hohenberg}(1993)}]{CrHo93}
\bibinfo{author}{\bibfnamefont{M.~C.} \bibnamefont{Cross}} \bibnamefont{and}
  \bibinfo{author}{\bibfnamefont{P.~C.} \bibnamefont{Hohenberg}},
  \bibinfo{journal}{Rev. Mod. Phys.} \textbf{\bibinfo{volume}{65}},
  \bibinfo{pages}{851} (\bibinfo{year}{1993}).

\bibitem[{\citenamefont{Nicolis}(1995)}]{Nico95}
\bibinfo{author}{\bibfnamefont{G.}~\bibnamefont{Nicolis}},
  \emph{\bibinfo{title}{Introduction to nonlinear science}}
  (\bibinfo{publisher}{Cambridge University Press},
  \bibinfo{address}{Cambridge}, \bibinfo{year}{1995}).

\bibitem[{\citenamefont{Thiele and Knobloch}(2006{\natexlab{b}})}]{dep1}
\bibinfo{author}{\bibfnamefont{U.}~\bibnamefont{Thiele}} \bibnamefont{and}
  \bibinfo{author}{\bibfnamefont{E.}~\bibnamefont{Knobloch}},
  \bibinfo{journal}{Phys. Rev. Lett.} \textbf{\bibinfo{volume}{97}},
  \bibinfo{pages}{204501} (\bibinfo{year}{2006}{\natexlab{b}}).

\bibitem[{\citenamefont{Savel'ev
  et~al.}(2004{\natexlab{b}})\citenamefont{Savel'ev, Marchesoni, and
  Nori}}]{sav04}
\bibinfo{author}{\bibfnamefont{S.}~\bibnamefont{Savel'ev}},
  \bibinfo{author}{\bibfnamefont{F.}~\bibnamefont{Marchesoni}},
  \bibnamefont{and} \bibinfo{author}{\bibfnamefont{F.}~\bibnamefont{Nori}},
  \bibinfo{journal}{Phys. Rev. Lett.} \textbf{\bibinfo{volume}{91}},
  \bibinfo{pages}{010601} (\bibinfo{year}{2004}{\natexlab{b}}).

\bibitem[{\citenamefont{Savel'ev et~al.}(2005)\citenamefont{Savel'ev,
  Marchesoni, and Nori}}]{sav05}
\bibinfo{author}{\bibfnamefont{S.}~\bibnamefont{Savel'ev}},
  \bibinfo{author}{\bibfnamefont{F.}~\bibnamefont{Marchesoni}},
  \bibnamefont{and} \bibinfo{author}{\bibfnamefont{F.}~\bibnamefont{Nori}},
  \bibinfo{journal}{Phys. Rev. E} \textbf{\bibinfo{volume}{71}},
  \bibinfo{pages}{011107} (\bibinfo{year}{2005}).

\bibitem[{\citenamefont{Doedel et~al.}(2001)\citenamefont{Doedel, Paffenroth,
  Champneys, Fairgrieve, Kuznetsov, Sandstede, and Wang}}]{AUTO}
\bibinfo{author}{\bibfnamefont{E.}~\bibnamefont{Doedel}},
  \bibinfo{author}{\bibfnamefont{R.}~\bibnamefont{Paffenroth}},
  \bibinfo{author}{\bibfnamefont{A.~R.} \bibnamefont{Champneys}},
  \bibinfo{author}{\bibfnamefont{T.~F.} \bibnamefont{Fairgrieve}},
  \bibinfo{author}{\bibfnamefont{Y.~A.} \bibnamefont{Kuznetsov}},
  \bibinfo{author}{\bibfnamefont{B.}~\bibnamefont{Sandstede}},
  \bibnamefont{and} \bibinfo{author}{\bibfnamefont{X.}~\bibnamefont{Wang}},
  \bibinfo{journal}{Technical Report, Caltech}  (\bibinfo{year}{2001}),
  \bibinfo{note}{url: http://cmvl.cs.concordia.ca/auto/}.

\bibitem[{\citenamefont{Risken}(1984)}]{risken}
\bibinfo{author}{\bibfnamefont{H.}~\bibnamefont{Risken}},
  \emph{\bibinfo{title}{The Fokker-Planck Equation}}
  (\bibinfo{publisher}{Springer, Berlin}, \bibinfo{year}{1984}).

\bibitem[{\citenamefont{Reimann}(2002)}]{reimann02}
\bibinfo{author}{\bibfnamefont{P.}~\bibnamefont{Reimann}},
  \bibinfo{journal}{Physics Rep.} \textbf{\bibinfo{volume}{361}},
  \bibinfo{pages}{57} (\bibinfo{year}{2002}).

\bibitem[{\citenamefont{Bordyugov and Engel}(2007)}]{bordyugov07}
\bibinfo{author}{\bibfnamefont{G.}~\bibnamefont{Bordyugov}} \bibnamefont{and}
  \bibinfo{author}{\bibfnamefont{H.}~\bibnamefont{Engel}},
  \bibinfo{journal}{Physica D} \textbf{\bibinfo{volume}{228}},
  \bibinfo{pages}{49} (\bibinfo{year}{2007}).

\bibitem[{\citenamefont{Strogatz}(1994)}]{Stro94}
\bibinfo{author}{\bibfnamefont{S.~H.} \bibnamefont{Strogatz}},
  \emph{\bibinfo{title}{Nonlinear Dynamics and Chaos}}
  (\bibinfo{publisher}{Addison-Wesley}, \bibinfo{year}{1994}).

\bibitem[{\citenamefont{Beltrame et~al.}(2011)\citenamefont{Beltrame, Knobloch,
  H{\"a}nggi, and Thiele}}]{BKHT11}
\bibinfo{author}{\bibfnamefont{P.}~\bibnamefont{Beltrame}},
  \bibinfo{author}{\bibfnamefont{E.}~\bibnamefont{Knobloch}},
  \bibinfo{author}{\bibfnamefont{P.}~\bibnamefont{H{\"a}nggi}},
  \bibnamefont{and} \bibinfo{author}{\bibfnamefont{U.}~\bibnamefont{Thiele}},
  \bibinfo{journal}{Phys. Rev. E} \textbf{\bibinfo{volume}{83}},
  \bibinfo{pages}{016305} (\bibinfo{year}{2011}).

\bibitem[{\citenamefont{Barker and Henderson}(1976)}]{barker76}
\bibinfo{author}{\bibfnamefont{J.~A.} \bibnamefont{Barker}} \bibnamefont{and}
  \bibinfo{author}{\bibfnamefont{D.}~\bibnamefont{Henderson}},
  \bibinfo{journal}{Rev. Mod. Phys.} \textbf{\bibinfo{volume}{48}},
  \bibinfo{pages}{587} (\bibinfo{year}{1976}).

\end{thebibliography}


\end{document}